\documentclass[12pt]{article}

\usepackage{amssymb,amsmath,amsfonts,authblk,booktabs,blindtext, eurosym,geometry,ulem,graphicx,caption,color,setspace,sectsty,comment,footmisc,caption,pdflscape,subfigure,array,hyperref,float, multirow, pbox, rotating, adjustbox, caption, regexpatch, siunitx, setspace,tabularx}
\usepackage[english]{babel}
\usepackage[flushleft]{threeparttable}
\usepackage{array}
\usepackage{ragged2e}
\usepackage{multicol}
\usepackage{verbatim} 
\usepackage{graphicx}
\usepackage{adjustbox}
\usepackage{tabularx}
\usepackage{xurl}
\usepackage{tablefootnote}
\usepackage[colorinlistoftodos]{todonotes}
\usepackage{atbegshi}
\usepackage{rotating}
\AtBeginDocument{\AtBeginShipoutNext{\AtBeginShipoutDiscard}}

\newcommand{\blue}[1]{{\color{blue} #1}}
\definecolor{darkblue}{rgb}{0.0,0.0,0.5}

\newcolumntype{P}[1]{>{\RaggedRight\hspace{0pt}}p{#1}}

\hypersetup{
	colorlinks= true,
	linkcolor = darkblue,
	urlcolor  = darkblue,
	citecolor = darkblue,
	anchorcolor = blue
}
\usepackage{csquotes}
\usepackage[style = apa, sorting = nyt,
			natbib = true,
			doi = false,
			isbn = false]{biblatex}
\begin{document} 

\title{The Marginal Effects of Ethereum Network MEV Transaction Reordering}
\author[1]{Bruce Mizrach \qquad Nathaniel Yoshida}\thanks{Correspondence: Department of Economics, Rutgers University, 75 Hamilton Street, New Brunswick, NJ 08901 USA. email:
mizrach@econ.rutgers.edu.  We would like to thank EigenPhi for providing the sandwich data and Peter Zimmerman for helpful comments.}
\affil[1]{\small Department of Economics, Rutgers University, New Brunswick, NJ USA}
\date{First Draft: August 2025 \\ Revised: September 2026}

\begin{titlepage}
\maketitle
\begin{abstract}
\noindent Two MEV builders now produce nearly 80\% of Ethereum blocks.  Block builders have the ability to reorder transactions on the blockchain in a way that can be harmful to participants.  We estimate participants would pay in the aggregate nearly \$7.2 million per month to guarantee that they remained in the first quartile of the block. Sandwich attacks, in which a transaction is front run, are frequent, averaging more than one every two blocks.  Gas fees on these transactions pay for nearly 9.6\% of the MEV payments to the validator.  Reforms such as gas fee priority or private transaction pools might be helpful.

\end{abstract}

\setcounter{page}{0}
\thispagestyle{empty}
\end{titlepage}


\pagebreak \newpage
\setstretch{1.05}
\section{Introduction} \label{introduction}

The Ethereum blockchain has evolved considerably since it was created in July 2015 by Vitalik Buterin.\footnote{\url{https://etherscan.io/block/1}}  Most tokens utilize the ERC-20 standard which was adopted in 2017. Since September 2022, blocks are validated by a system called proof-of-stake which replaced the energy intensive \textit{mining} in proof-of-work.  Validators who stake Ether, the network's utility token, are eligible to approve blocks.  Block creators receive rewards in the form of \textit{gas} fees, payments denominated in Ether to defer the processing and storage costs of adding new blocks to the chain.

Buterin raised concerns about the centralization of the network under proof-of-stake.  Validators have twelve seconds to approve a block with hundreds of transactions, which can occur at any time in the day.  These high fixed costs and technical sophistication might lead to a small group of staking pools validating most of the blocks.  In the first month after the merge, \citet{KapengutPoS} found that the Ethereum network Herfindahl index rose slightly compared to the mining pools that were validating blocks under proof-of-work.

\citet{ButerinPBS} advanced the idea of \textit{proposer builder separation} as a potential solution.  In this framework, the validator can choose from a set of proposed builder blocks.  The block builder seeks to \textit{maximize extractable value} (MEV) in the block by reordering transactions or inserting transactions from \textit{searchers.\footnote{Searchers are participants who scan the public mempool for profitable arbitrages}}  This extractable value was previously obtained by the miner under proof-of-work.  From the perspective of the Ethereum developers this proposer-builder separation is a way for validators to share in MEV revenue and enable less sophisticated validators to participate in the blockchain.  EigenPhi estimates\footnote{https://eigenphi.substack.com/p/30m-72-of-searchers-mev-revenue-went} that in January-February 2023, 72\% of MEV extracted went to validators. 

Toni Wahrstätter, a researcher at the Ethereum Foundation, has created a dashboard\footnote{\url{MEVBoost.pics}} for the number of blocks built by maximum-extractable value (MEV) algorithms.  On September 23, 2022, shortly after the Beacon merge, MEV builders completed 34.2\% of blocks. By December 2022, the MEV share had grown to over 90\%.

\begin{figure}[H]
	\centering
		\caption{MEV Block Formation}
		\label{fig:mev_share}
        \begin{minipage}{0.97\linewidth}
        \begin{center}
			\includegraphics[width=0.97\textwidth]{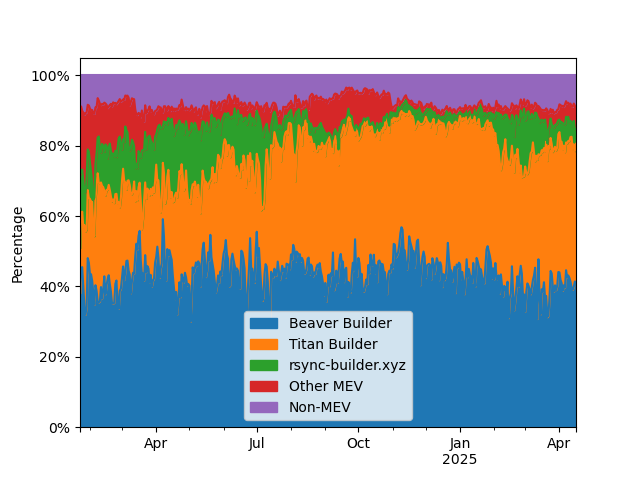} \\
		\end{center}
		\small
		\texttt{Note: } The MEV builder addresses are from MEV Boost.pics, and we detect payments in Mainnet blocks from them to the validators. The sample is from January 23, 2024 to April 16, 2025.
    \end{minipage}
\end{figure}

While validators are still chosen randomly, many use open source software from Flashbots called MEV-Boost\footnote{https://docs.flashbots.net/flashbots-mev-boost/introduction} to access a selection of proposed blocks. They generally select the block offering the highest payment to the validator.  We typically observe the payment of ETH to the validator as the last transaction in the block.  Information on the builders and relays is from Mevboost.pics. From these transactions, sourced from Google Big Query,\footnote{bigquery-public-data.crypto\_ethereum}  we can compute the market share of the MEV builders.  By August 2024, as shown in Figure \ref{fig:mev_share}, just two builders have come to dominate the market.

Some MEV arises because of efficiencies in gas utilization.  The Ethereum foundation notes\footnote{ These practices are known as \textit{gas golfing}, \url{https://ethereum.org/en/developers/docs/mev/}} two examples:
``[1] using addresses that start with a long string of zeroes ...since they take less space (and hence gas) to store; [2] leaving small ERC-20 token balances in contracts, since it costs more gas to initialize a storage slot... than to update a storage slot.'' 

The other sources of MEV are less benign.  They include various types of profitable manipulation of the transactions in the block.  \citet{EskandariFront} describes several categories of front running behavior.  They include \textit{displacement} in which a builder simply substitutes their own address for a potentially profitable transaction. A second group involves the insertion of transactions, generally \textit{sandwiching} an existing mempool transaction between two new transactions.  A third category is suppression of a transaction to a later sequence in the block or even to a subsequent block.  \citet{CapponiFrontRun} estimate that of the 6.6 million blocks with transactions on Uniswap V2 decentralized exchanges between May 15, 2020 and January 13, 2024, more than 90\% were at risk of front running, and more than one million blocks actually experienced front running attacks.

\begin{figure}[H]
	\centering
		\caption{Validator Revenue}
		\label{fig:validator_revenue}
        \begin{minipage}{0.97\linewidth}
        \begin{center}
			\includegraphics[width=0.97\textwidth]{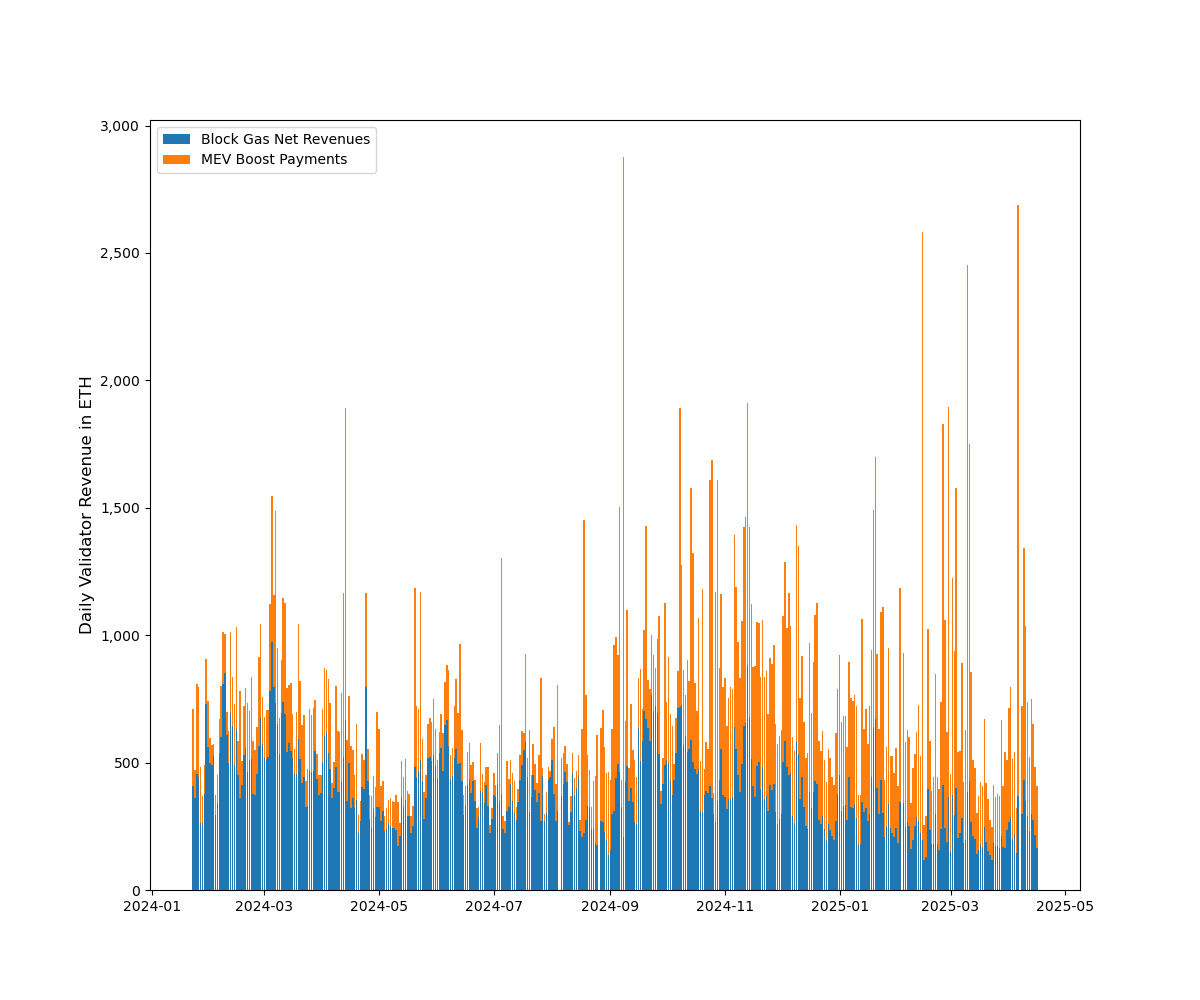} \\
		\end{center}
		\small
		\texttt{Note: } The MEV payments data are from the last transaction in the block and range from January 23, 2024 to April 16, 2025.  The net gas revenue calculations utilize Ethereum Mainnet blocks sourced from Google Big Query.  We exclude four surges in MEV payments: August 5, 2024 that coincides with a 30\% price decline in ETH that appears to have multiple causes\footnote{\url{https://medium.com/@bikingex/crypto-market-black-monday-on-august-5th-2024-9a2c0dbf7430}}; August 26, 2024 when the Ethereum Foundation transferred a large supply of ETH to Kraken.\footnote{\url{https://blockcast.cc/market-roundup-bitcoin-and-ethereum-26-august-2024/}}February 3, 2025 in which Ethereum suffered a 27\% flash crash\footnote{\url{https://www.soliduslabs.com/post/ether-feb3-flash-crash-a-stark-reminder-of-crypto-market-vulnerabilities}}; and April 7, 2025 is the day that U.S. President Trump announced new global tariffs.
    \end{minipage}
\end{figure}

We show in Figure \ref{fig:validator_revenue} that MEV Boost revenues are substantial.  They comprise gas fees net of base fees that are burned, and payments from MEV algorithms.

Median total revenues are 635 ETH per day (\$1.59 million per day at an ETH price of \$2,500), with MEV payments making up 41\% of that total. 

There is an active literature within the Ethereum community to address protocol changes that might be the fairest for validators and/or reduce the harmful effects of front running.  Several proposals have been offered to smooth out MEV payments across validators, including full sharing of MEV payments,\footnote{\url{https://ethresear.ch/t/committee-driven-mev-smoothing/10408}} burning MEV,\footnote{\url{https://ethresear.ch/t/mev-burn-a-simple-design/15590}} offering revert protection to failed transactions (\cite{zhu_revert}), and more complex auction designs (\cite{monnot_auctions}). To our knowledge, the community has not proposed a system of absolute time priority.  This is particularly surprising since Ethereum developers acknowledge that there is an incentive to bid as late as possible in an MEV auction.\footnote{\url{https://ethresear.ch/t/reducing-latency-games-by-levelling-the-playing-field-on-block-size-for-pbs/19356}}

We estimate the harm to participants by estimating the probability that a transaction will be included in the first group of block transactions. This can be thought of as analogous to a trader paying to be first in the queue for a news driven trading strategy. We provide an ex ante measure of the gas that market participants would have paid  to be included in this group. Throughout the paper, we call this the \textit{shadow price of time priority}.

We utilize a logit model for the probability that a transaction will be placed in the first quartile of the completed block. Using these estimates, we compute the marginal effects from transactions that the MEV agent is most likely to move into favorable positions.  These marginal effects can be converted into the gas units that the participant would have needed to provide to ensure a specific position in the finalized block. This is paramount to undoing the harmful reordering of transactions, deviations in the execution order of transactions between pending transactions in the mempool, and the execution order of those transactions in the finalized block. We find that these costs are statistically and economically significant.  Ethereum network participants would need to pay an average of \$0.21 more per transaction or a total of nearly \$7.2 million per month to undo the effects of the MEV agent.

These costs are especially high for sandwich transactions.  We find that the front run transaction in a sandwich attack is 48\% more likely to appear in the first block quartile. This estimate is independent of poor execution quality due to being front run.
The back run gas fees of a sandwich provide approximately 9.6\% of the total fee paid by the block builder to the validator.  

Our path forward is to quantify the dominance of MEV agents in block formation on the Ethereum network in Section \ref{network concentration}. Section \ref{Approaches to Transaction Ordering} contrasts financial market approaches to transaction ordering to those on the blockchain.  Section \ref{Baseline Specification} contains our baseline model for the probability of reordering blockchain transactions from their prior order in the mempool.  This model provides marginal effects of the harm to participants.  Section \ref{sandwich} discusses sandwich attacks.  These attacks are numerous, averaging more than one every two blocks, and the marginal effects of the sandwich attacks are large compared to those we find in the baseline model.  

We turn now to a largely unexpected consequence of the move to proof-of-stake, the dominance of a handful of MEV builders in forming Ethereum blocks.

\section{Network Concentration} \label{network concentration}

We computed the number of blocks and MEV extracted for all builders from January 23, 2024 to April 16, 2025.  The top 10 builders ranked by MEV can be seen in Table \ref{tab:builder_share}.

\begin{table}[H]
\centering
\begin{threeparttable}
  \caption{Block Builder Market Shares}
  \label{tab:builder_share}%
\begin{tabular}{lrr}
\hline
Builder & MEV Value (ETH) & Block Count \\
\hline \hline
beaverbuild.org & 111907.12 & 1,390,451 \\
Titan Builder & 77,989.67 & 1,070,749 \\
rsync-builder.xyz & 31,392.580577 & 271,284 \\
NotMEV & 0.000000 & 265,409 \\
Ultrasound Money & 2,439.68 & 3,7530 \\
Flashbots & 5,186.19 & 34,316 \\
jetbldr.xyz & 916.93 & 23,614 \\
@penguinbuild.org & 695.03 & 9,457 \\
Builder+ www.btcs.com/builder & 1,106.77 & 7,742 \\
f1b.io & 591.957 & 7,050 \\
\hline
\end{tabular}
     \begin{tablenotes}
      \small
      \item \texttt{Note: } The sample is January 23, 2024 to April 16, 2025. The block counts, builder mapping, and MEV values are from mevboost.pics.
    \end{tablenotes}
    \end{threeparttable}%
\end{table}

The dominant block builders are beaverbuild.org,\footnote{\url{https://beaverbuild.org/} as of May 6, 2025, Beaverbuild has rebranded as BuilderNet \url{https://buildernet.org/blog/beaverbuild}} Titan Builder,\footnote{\url{https://www.titanbuilder.xyz/}} and  rsync-builder. Beaverbuild provides very little documentation apart from a picture of a beaver and their motto:  ``be mergin, be splurgin and by god be searchin."  Titan Builder emphasizes that they are neutral, meaning they allow their group of searchers to propose a variety of potential blocks.  Rsynch is as minimalist as beaverbuild, only providing a series of endpoints for block proposals.

Prior to the Beacon merge, blocks were added by the miners. For the month leading up to the merge, August 14 to September 14, 2022, \citet{KapengutPoS} estimated the Herfindahl miner index at 1,245.  Ethermine had the largest market share at 28.6\%.  In the month following the merge, from September 16 to October 16, 2022, the Herfindahl index for the top 10 validators was 1,009, a 19\% decline in concentration.

91.1\% of all blocks are completed by MEV agents.  As for market share, the top three builders have 85.9\% of the market share of all blocks, and 86.8\% of the share of MEV.  The Herfindahl index for the share of blocks is 3,186.  This is 2.5 times more than the concentration of miners before the merge and a 200\% increase from the month after the merge. Our empirical findings of increased network and MEV concentration coincide with the predictions of the theoretical model of \citet{Burian_MEV}. They propose a further separation of payload proposal following an execution tickets model. They note that when buyers are heterogeneous, execution tickets should become concentrated among buyers with low capital costs and high MEV extraction capability. 

The growing centralization of the Ethereum network remains a challenge for its developers.  Our concerns go beyond that to look at the harms caused by the lack of time priority.

\section{Approaches to Transaction Ordering}\label{Approaches to Transaction Ordering}

\subsection{Time Priority in Equity Markets}

In  2004, all of the major market making firms on the New York Stock Exchange (NYSE) were fined\footnote{https://www.sec.gov/news/press/2004-42.htm} a total of more than \$240 million dollars for trading ``...their dealer accounts ahead of executable agency orders on the same side of the market, orders that were executed later at prices inferior to the prices of dealer account trades. At other times, the specialists traded ahead of executable limit orders, which then went unexecuted and ultimately were canceled by the customers entering the orders.''

Since 2005, the U.S. equity market has been subject to the SEC order protection Rule 611\footnote{SEC Rule 611 of the National Market System, \url{https://www.federalregister.gov/documents/2005/06/29/05-11802/regulation-nms}} that prioritizes orders on the basis of price and time.  Rule 611 does not explicitly require price-time priority, but nearly every exchange and trading center has adopted these rules.\footnote{One exception is the CBOE Edge X which gives priority to retail traders, \url{https://www.cboe.com/us/equities/trading/offerings/retail_priority/}. The NYSE parity/allocation model \url{https://beta.nyse.com/publicdocs/NYSE_broker_systems_and_parity_priority_allocation_model.pdf} gives priority to NYSE floor brokers and market makers.  \citet{BattalioNYSE} estimate that in 2017, this program provided a subsidy of more than \$9 million to NYSE floor participants.} Given the emphasis on time priority, trading firms have built a low latency infrastructure to try to execute faster. These investments include co-location\footnote{See for example on the Nasdaq, \url{https://www.nasdaq.com/solutions/nasdaq-co-location} and the NYSE \url{https://www.theice.com/data-services/global-network }} of hardware and software directly at the exchanges as wells as the construction of microwave and fiber optic networks (\cite{shkilko_micro} \& \cite{euro_Latency}) to accelerate news trading and to arbitrage price discrepancies across trading centers.

\subsection{Blockchain Mempools}
Blockchains have so far not come under direct regulatory supervision by the federal government.  Blockchains propose their own rules, and there are mechanisms for comment and approval by the stakeholders in the blockchain.  Ethereum governance takes place off-chain.  Proposals are distributed, commented upon, and then approved by a set of core developers.\footnote{\url{https://ethereum.org/en/governance/}}  The Ethereum Foundation clearly recognizes MEV as a problem, but none of the potential reforms, e.g. EIP-7732\footnote{https://eips.ethereum.org/EIPS/eip-7732} which decouples execution and consensus validation, have been adopted.

The Ethereum network infrastructure is by design more decentralized.  Any user can run a node, and the clients, such as Geth, contain a transaction pool (mempool) which will distribute pending transactions from other nodes. In a sample of over 35 million transactions in our data set, we find that on average just over 98\% of finalized block transactions are found within the public mempool.\footnote{\cite{Capponi_JFE} find private pool adoption prior to the merge stabilized at nearly 50\%.} 

We rely on mempool data from BlockNative through January 2025,\footnote{\url{https://www.blocknative.com/ethernow-sunset}} and from February 2025, ethPandaOps.\footnote{\url{https://ethpandaops.io/data/xatu/schema/mempool_/}}  Both sources provide Mainnet transaction hashes and timestamps for the arrival of the transaction in different geographic mempools.

Our data logs all mempool transactions across nodes in multiple geographical regions for the Ethereum mainnet blockchain. This includes information on when transactions entered, exited, replaced, finalized, or were rejected.  We report the first appearance in the mempool data for determining time priority.\blue{\footnote{This follows the methodology used in the empirical analysis of \citet{Adams_Slippage}. Moreover, we believe this timing is theoretically correct as it aligns to when the transaction was first broadcast to the other market participants.}}

\section{Baseline Specification} \label{Baseline Specification}

\subsection{Block position} \label{Mempool}

We compute two measures of block position, the position on the [0,1] interval of the first arrival time of transactions to the mempool, and the ex-post position in the fully constructed block, again normalized to [0,1].  We group the transactions into mempool and block quartiles for noise reduction and estimate the model daily based on a UTC clock.

Addresses to identify MEV builders are from MEVBoost.pics.  Addresses for centralized exchanges (CEX) and decentralized exchanges (DEX) are from Brian Lect's Github,\footnote{\url{https://github.com/brianleect/etherscan-labels/blob/main/data/etherscan/combined/combinedAccountLabels.json}} and all active addresses were confirmed manually from Etherscan. DEX are on-chain peer-to-peer marketplaces where users can swap a wide variety of tokens.  The largest Ethereum spot DEXs are Uniswap and Curve.\footnote{https://coinmarketcap.com/rankings/exchanges/dex/?type=spot}   Centralized exchanges are off-chain and users there can often trade much faster and exchange their tokens for fiat currency.  The largest spot CEXs include Binance, ByBit, and Coinbase.\footnote{https://coinmarketcap.com/rankings/exchanges/} 

\subsection{Factors impacting block position} \label{block position}

We consider an ordered logit model of the form
\begin{equation} \label{eq: 1}
    \mathbb{P}(\text{block quartile} = j | X) = F(\beta^{\top}X), \text{ } j \in \{1,2,3,4\}, \\
\end{equation}
where $F$ is the logistic distribution function.\footnote{We also considered an ordered probit model corresponding to F being the normal distribution function. Qualitatively, the results from probit and logit are nearly identical. See the appendix for summary tables and average marginal effects.} $X$ contains a measure of the maximum fee in gas units a participant will commit to the transaction. This encompasses both standard gas fees for the underlying transaction and a potential priority fee, directly rewarding the block builder for including the transaction in the final block. $X$ also contains three dummy variables corresponding to a transaction’s position in the mempool and four dummy variables denoting a transaction going to and from a DEX and to and from an MEV agent. We further control for the average gas price per block, the gas used for each transaction, and the payment from the builder to the validator. The average gas price per block serves as a measure of network congestion. The gas used in each transaction reflects the computational time of the transaction. Finally, the payment from the builder to the validator accounts for the direct economic incentive for validators to allow for transaction reordering to occur.\footnote{To address numerical instability in the Hessian matrix and to make computation more efficient, we first applied z-score normalization to all continuous variables. We then undo the scaling when reporting summary statistics and marginal effects by dividing by standard deviations.} 

The marginal effects of the model defined in Equation (\ref{eq: 1}) measure the change in probability of a transaction appearing in the first quartile. Mathematically, marginal effects are defined as partial derivatives of the distribution function.

\begin{equation} \label{eq: 2}
        \text{Marginal Effect}_i = \frac{\partial F}{\partial x_i} \\
\end{equation} 

If no transaction reordering was occurring, the only significant indicator of a transaction's position in the finalized block would be its position in the mempool. If mempool position is not a significant predictor of finalized block position but only max fee per gas is, then there would be evidence of reordering but the mechanism of reordering would be transparent and fair. Those with the highest valuation for rapid transaction execution would pay the most. Neither of these scenarios is consistent with our empirical findings.

We will later report estimates for the entire month of October 2024, but we illustrate the model's estimation by reporting results in Table \ref{tab:ordered_model_estimates_probit} using transaction data from October 1, 2024.\footnote{We show in the appendix that results that are significant under iid standard errors remain significant under robust standard errors.}\textsuperscript{,}\footnote{Using the test methodology developed in  \citet{Wooldridge_2014} and the average gas price per block one block and two blocks back as instruments, we did not find evidence of endogeneity for the max fee per gas at a 5\% level. To be conservative, we include average gas price as a control variable.}

\begin{table}[H]
\centering
\begin{threeparttable}
\caption{Model Estimates October 1, 2024}
\label{tab:ordered_model_estimates_probit}
\begin{tabular}{lrr}
\toprule
Variable & Value & Standard Error \\
\midrule
\textit{Main Variables}\\
max fee per gas & ${-1.291 \times 10^{-3}}^{***}$ & $3.294 \times 10^{-5}$  \\
to DEX & ${-3.826 \times 10^{-1}}^{***}$ & $1.466 \times 10^{-2}$ \\
to MEV & ${-1.597}^{***}$ & $2.990 \times 10^{-2}$ \\
from DEX & ${-1.599}^{***}$ & $8.503 \times 10^{-2}$ \\
from MEV & ${1.679}^{***}$ & $3.259 \times 10^{-2}$ \\
\textit{Mempool Position} \\
mempool quartile 1 & $-7.709 \times 10^{-3}$ & $1.075 \times 10^{-2}$ \\
mempool quartile 2 & $3.520 \times 10^{-3}$ & $1.079 \times 10^{-2}$ \\
mempool quartile 3 & $-6.216 \times 10^{-3}$ & $1.079 \times 10^{-2}$ \\
\textit{Controls}\\
average price & ${8.322 \times 10^{-4}}^{***}$ & $1.854 \times 10^{-4}$\\
gas used & ${-4.179 \times 10^{-6}}^{***}$ & $4.702 \times 10^{-8}$\\
validator payment & ${9.143 \times 10^{-5}}^{*}$ & $4.253 \times 10^{-5}$\\
Boundary 1|2 & ${-6.733 \times 10^{-1}}^{***}$ & $8.175 \times 10^{-3}$  \\
Boundary 2|3 & ${3.127 \times 10^{-1}}^{***}$ & $8.085 \times 10^{-3}$  \\
Boundary 3|4 & ${1.189}^{***}$ & $8.469 \times 10^{-3}$  \\
\bottomrule
\end{tabular}
\begin{tablenotes}
\small
\item \textit{Note:} Negative coefficients indicate a higher likelihood of a transaction being placed earlier in the block. $***$ indicates significance at the 0.1\% level. $*$ indicates significance at the 5\% level. 
    
\end{tablenotes}
\end{threeparttable}
\end{table}

The mempool position of a transaction is not a statistically significant predictor of its position in the block. This demonstrates that block builders are not preserving time priority in their construction of blocks. The maximum fee per gas, going to DEX, going to an MEV agent, coming from DEX, and coming from an MEV agent are all statistically significant predictors of a transaction's block position.  DEX transactions are much more likely to move up, whether they are transactions to or from the DEX. The largest effect is whether the transaction involves the MEV.\footnote{In the appendix, we report results of a 5-category and 10-category model. Results remain consistent in terms of sign and significance. However, computational time increases significantly.}

The average marginal effects for the main variables of interest are in Table \ref{tab:avg_marginal_probit}. These marginal effects reflect the full model with controls.

\begin{table}[H]
  \centering
  \begin{threeparttable}
  \caption{Average Marginal Effects for October 1, 2024}
    \begin{tabular}{lrrr}
    \toprule
          & & Marginal Effects        &  \\
    Variable & \multicolumn{1}{l}{Prob. In First Quartile} & \multicolumn{1}{l}{Gas Equivalent Cost} & \multicolumn{1}{l}{Cost in USD} \\
    \midrule
    \midrule
    $\text{max fee per gas}^*$ & 1\% & 36.74 & \$0.0021 \\
    to DEX & 8.42\% & 310   & \$0.0180 \\
    to MEV & 35.79\% & 1,316 & \$0.0767 \\
    from DEX & 35.46\% & 1,303 & \$0.0760 \\
    from MEV & -25.25\% & 928 & \$0.0541 \\
    \bottomrule
    \end{tabular}%
  \label{tab:avg_marginal_probit}%
     \begin{tablenotes}
      \small
      \item \texttt{Note: } $*$ denotes that the variable max fee per gas is continuous and thus we normalize effects to be equivalent to a 1\% change in the probability of appearing in the first block quartile. All other variables are discrete and thus marginal effects reflect the change in probability when the value of the variable goes from 0 to 1. The table provides estimates of the marginal effects as the change in the probability of the transaction being included in the first quartile.  We measure the harm to the participant by measuring the gas and dollar cost of achieving a similar block position.  We use average daily gas costs of 22.45 Gwei which we obtain from Etherscan and the closing Ethereum price of \$2,597.34 from CoinGecko.
    \end{tablenotes}
    \end{threeparttable}%
\end{table}%
We can transform the marginal effects of the transaction characteristic variables into gas equivalents by scaling by the marginal effect of the max fee per gas,

\begin{equation} \label{Eq: 3}
        \text{Marginal Effect in Gas Units}_i = \frac{\partial F/\partial x_i}{\partial F/\partial \text{max fee per gas}}. \\
\end{equation} 

A transaction to DEX is approximately 8\% more likely to be in the first quartile of the block, all else held constant, with similar interpretations for the other marginal effects. Equivalently, approximately 37 additional units of gas would be required to increase the probability that a transaction appears in the first quartile of block transactions by 1\%.   As an additional point of comparison, the average maximum fee per gas on October 1st was just under 70 units of gas. At an average daily gas price of 22.45 Gwei per unit of gas and a closing price of \$2,597.34 per ETH, a 1\% increase in the probability of a transaction being in the first block quartile costs approximately \$0.002. 

Marginal effects on par with those for the dummy variables to DEX, to MEV, from MEV, and from DEX cost between 310 and 1,316 additional units of gas. Converting this to US dollars, such marginal effects cost between \$0.0180 and \$0.0767 of additional gas. 

As previously discussed, we provide an ex ante measure of the premium market participants are willing to pay to have their order placed first. We limit possible actions to what market participants can accomplish in 12 seconds. This rules out transferring assets to other blockchains. We take market participants to be price takers with regard to gas fees which are known to be highly persistent.

It should be noted that this does not provide guaranteed protection against sandwich attacks as discussed in Sections \ref{sandwich} and \ref{sandwich effects}. It should also be noted that the only way to ensure the most favorable execution order is to purchase the block. Many market participants, however lack the sophistication and resources to construct blocks. Therefore, while paying a premium to ensure placement among the highest priority group does not entirely protect a market participant from MEV, it is the only countermeasure that is accessible to all market participants.

By purchasing the block, the block builder is afforded complete control of the transaction order within the block, including the insertion of transactions from searchers that were not previously in the block. We consider the gas required to ensure that a transaction appears in the first quartile of block transactions. This is the shadow price of time priority. Specifically, we consider how much additional gas is required to ensure a desired block position for a transaction within the finalized block. Thus, with sufficient additional gas, there is enough of an economic incentive such that the validator is better off with transaction order in the finalized block being consistent with transaction order in the mempool, i.e., preserving time priority in transaction execution. 

For October 1st, we find an average of 3,674 additional units of gas provide a 100\%  assurance of inclusion in the high priority group. This is equivalent to approximately \$0.21 of additional gas per transaction. As every transaction is subject to reordering once the block has been purchased, we aggregate over all daily transactions. There were 1,266,656 transactions recorded for October 1st. Thus, the shadow price of time priority is approximately 4.654 billion units of gas. This is equivalent to almost \$272,000 of gas based on an average daily gas price of 22.45 Gwei and a closing Ethereum price of \$2,597.34.

\begin{figure}[H]
    \centering
    \caption{Daily Average Marginal Effects}
    \label{fig:daily_AME}%
        \begin{minipage}{0.97\linewidth}
        \begin{center}
		  \begin{tabular}{cc}
         \includegraphics[width = 0.4\textwidth]{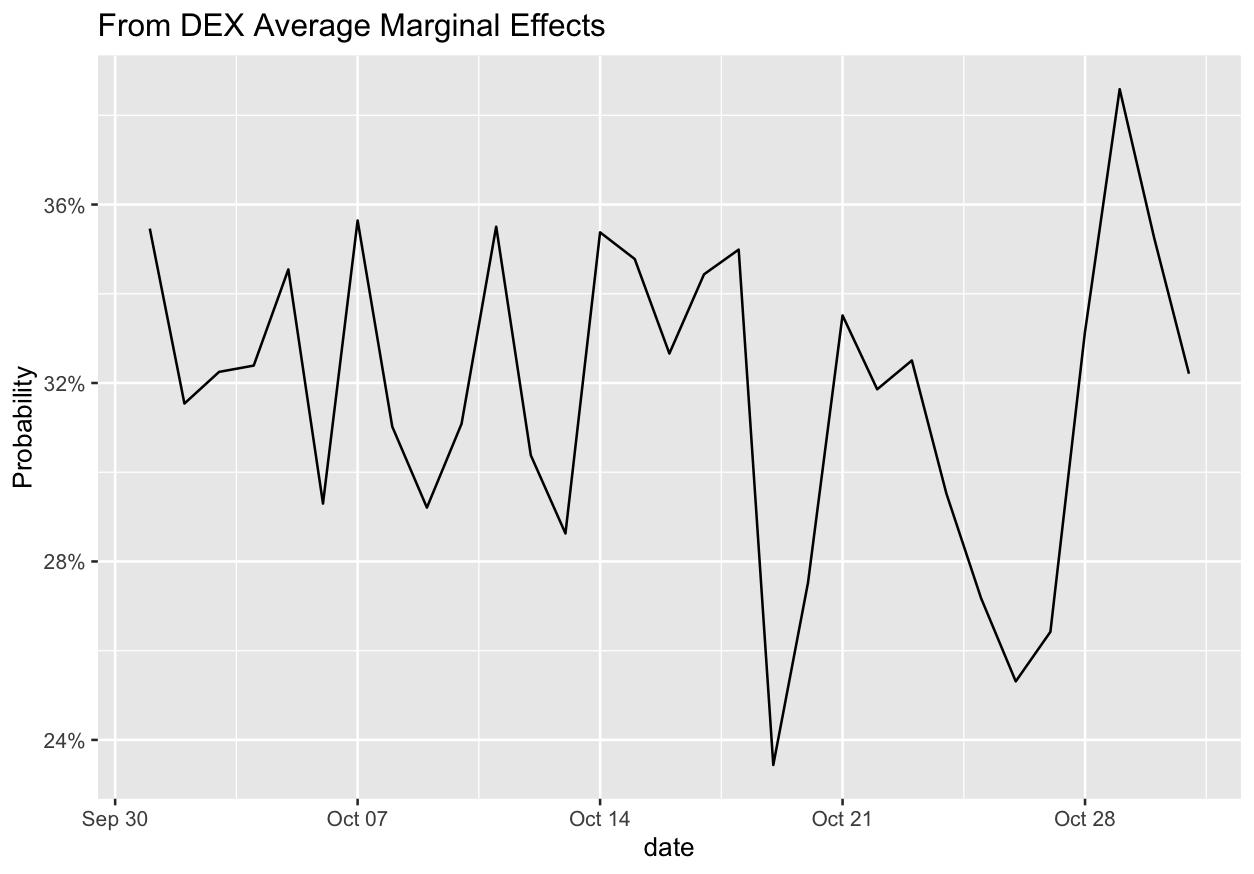} & \includegraphics[width = 0.4\textwidth]{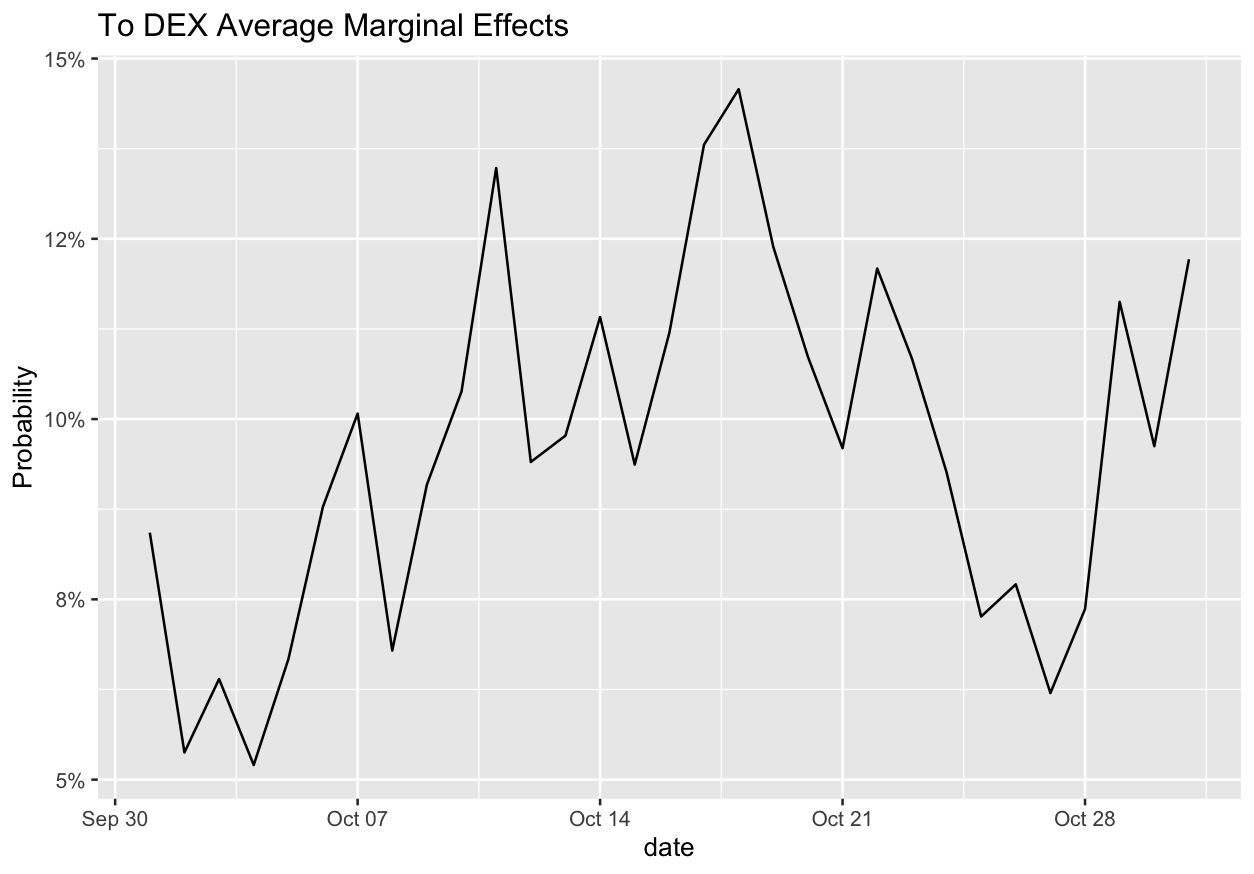} \\
         \includegraphics[width = 0.4\textwidth]{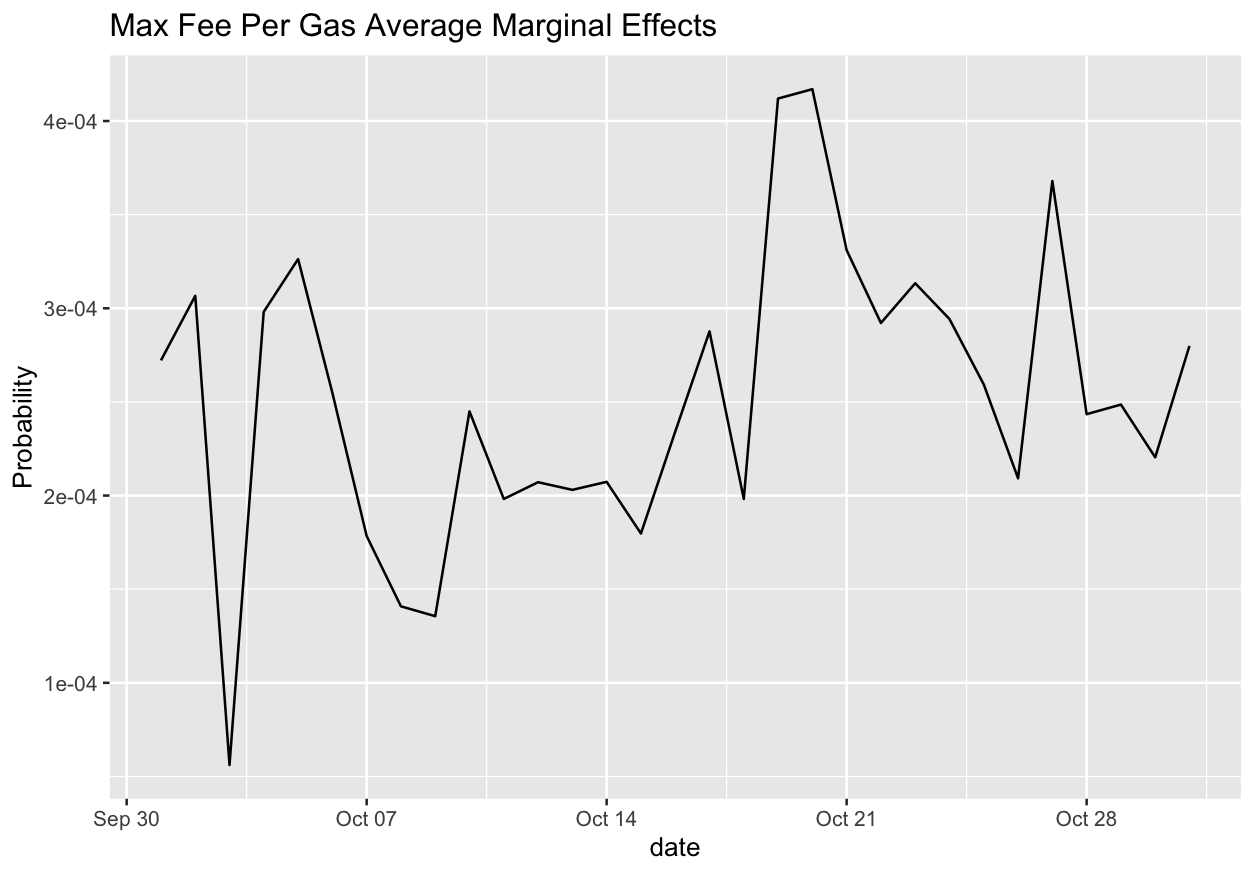} & \includegraphics[width = 0.4\textwidth]{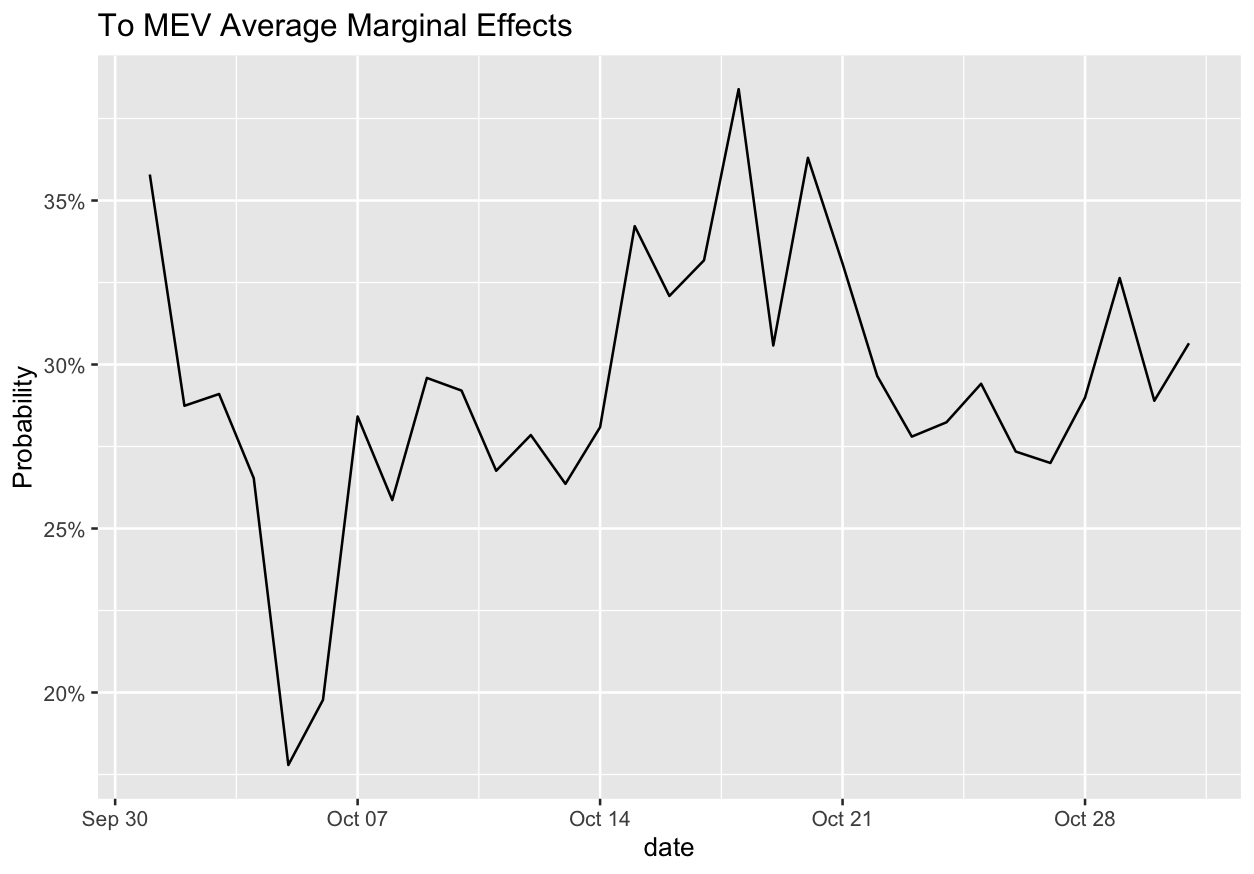} \\ 
    \end{tabular}	
		\end{center}
		\small
		\texttt{Note: } Average marginal effects correspond to increased (decreased) probability of a transaction being in the first quartile of block transactions
        \end{minipage}
    \label{fig:enter-label}
\end{figure}

We now re-estimate the model for each day in our sample. We average marginal effects within each day in our sample and graph our results in Figure \ref{fig:daily_AME}.

The average marginal effects for each day between October 1st and October 31st are shown in Figure \ref{fig:daily_AME}. With the exception of the marginal effects from maximum fee per gas, the marginal effects remain relatively stable and close to those calculated for October 1st throughout the period. Thus, we see that the transaction characteristics measured by the dummy variables from DEX, to DEX, and to MEV have large and relatively stable impacts on the position of the transaction within the final block. 

Although the marginal effects for max fee per gas are orders of magnitude smaller than the marginal effects for the other regressors of interest, significant day-to-day changes in this estimated marginal effect are evident. Due to the large day-to-day changes in the marginal effect for maximum fee per gas, the economic costs can vary greatly. In combination with large daily movements in the price of ETH and gas prices, the economic effects can vary considerably. The lowest marginal effect observed for maximum fee per gas is approximately $0.0056\%$ measured on October 3rd. The highest marginal effect observed for maximum fee per gas is $0.0417\%$ observed on October 20th. This implies that increasing the probability of a transaction appearing in the first block quartile by 1\% costs between 24 and 179 additional units of gas. 

\begin{figure}[H]
    \centering
    \caption{Daily Inclusion Costs}
    \label{fig:daily_reordering_costs}
        \begin{minipage}{0.97\linewidth}
        \begin{center}
		  \begin{tabular}{cc}
         \includegraphics[width = 0.4\textwidth]{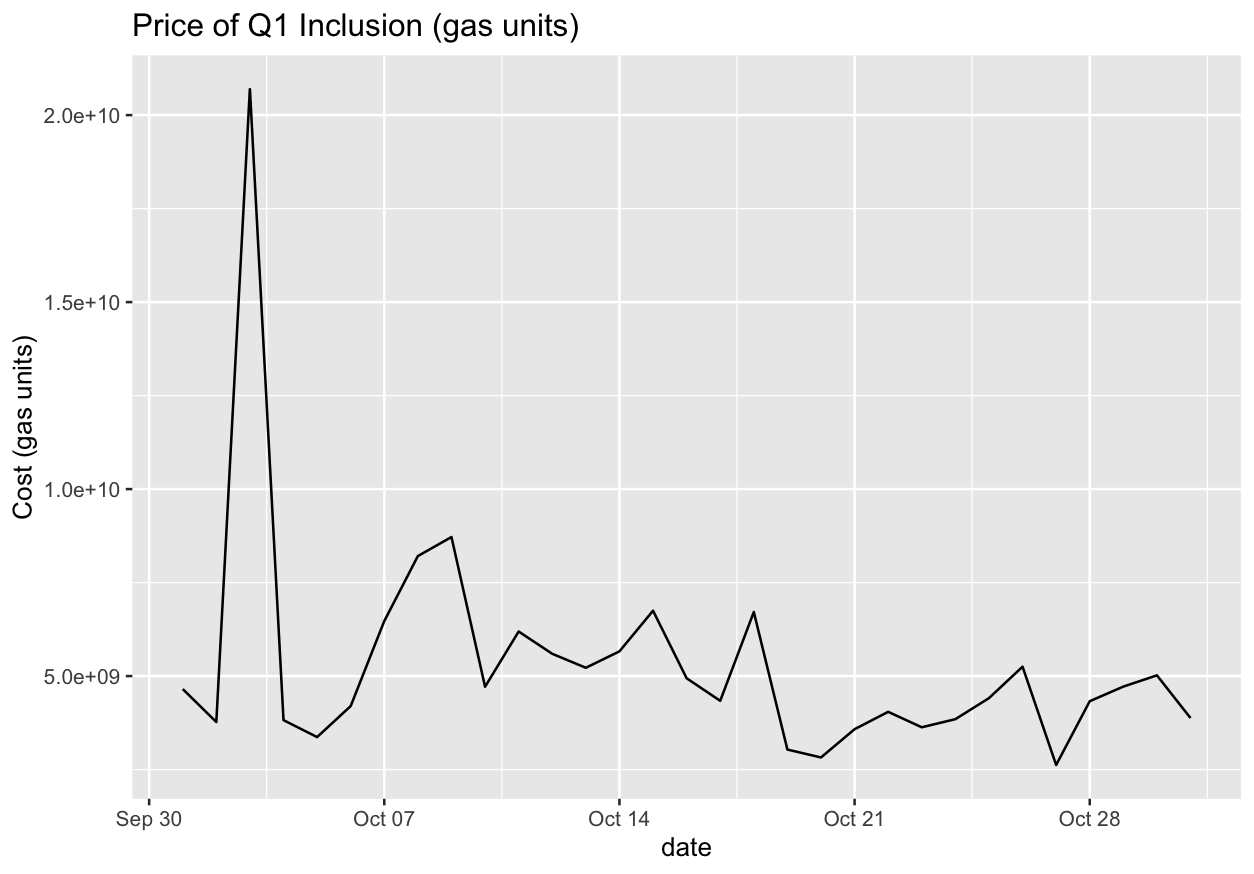} & \includegraphics[width = 0.4\textwidth]{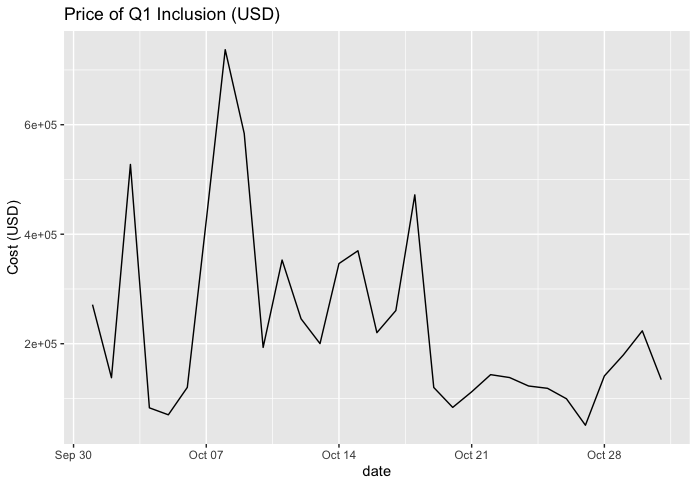} \\
    \end{tabular}	
		\end{center}
		\small
		\texttt{Note: } The graph on the left shows the shadow price of time priority, aggregated across all daily transactions, in units of gas, while the right graph shows the equivalent cost in USD.  
        \end{minipage}
    \label{fig:enter-label}
\end{figure}

The daily shadow price of time priority, aggregated across all daily transactions, is shown in Figure \ref{fig:daily_reordering_costs}. We find that approximately 4.818 billion units of additional gas per day on average are needed to prevent transaction reordering. This is equivalent to just over \$225,000 of gas per day on average. The most expensive day to ensure time priority in terms of gas units was October 8th, when it cost about 8.209 billion units of additional gas or about \$737,000 of additional gas. The cheapest day to ensure time priority was October 27th, when approximately 2.623 billion units of additional gas were required. This is equivalent to approximately \$51,000 of gas. This shows both the large, economically significant losses to investors as a result of transaction reordering and the large variability in investor losses due to day-to-day changes in the marginal effects, as well as the large day-to-day changes in gas and Ethereum prices.  

\subsection{Quantile Marginal Effects} 
\label{quantile_ME}
We now turn to the analysis of the distribution of marginal effects.

\begin{figure}[H]
    \centering
    \caption{Daily Quantile Marginal Effects}
    \label{fig:daily_quantile}
        \begin{minipage}{0.97\linewidth}
        \begin{center}
		  \begin{tabular}{cc}
         \includegraphics[width = 0.4\textwidth]{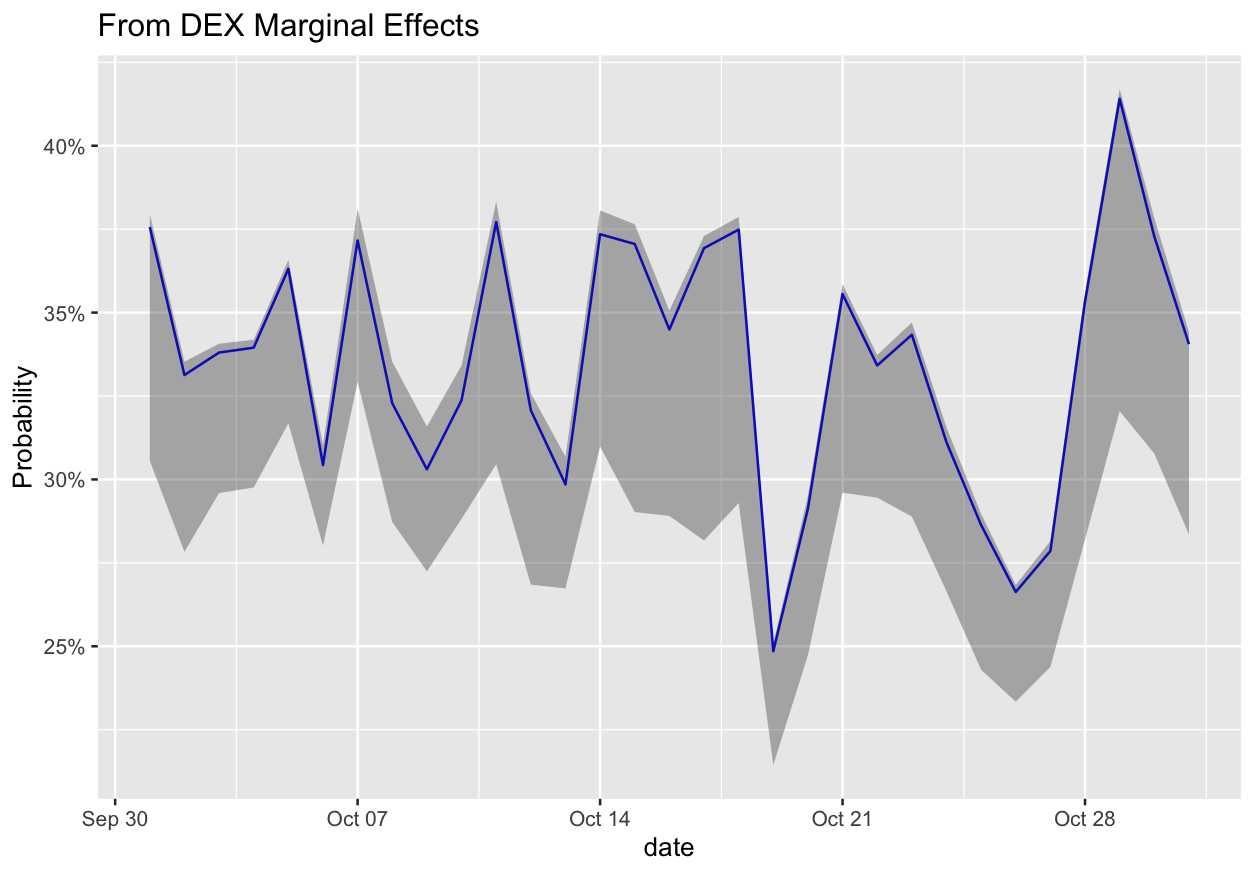} & \includegraphics[width = 0.4\textwidth]{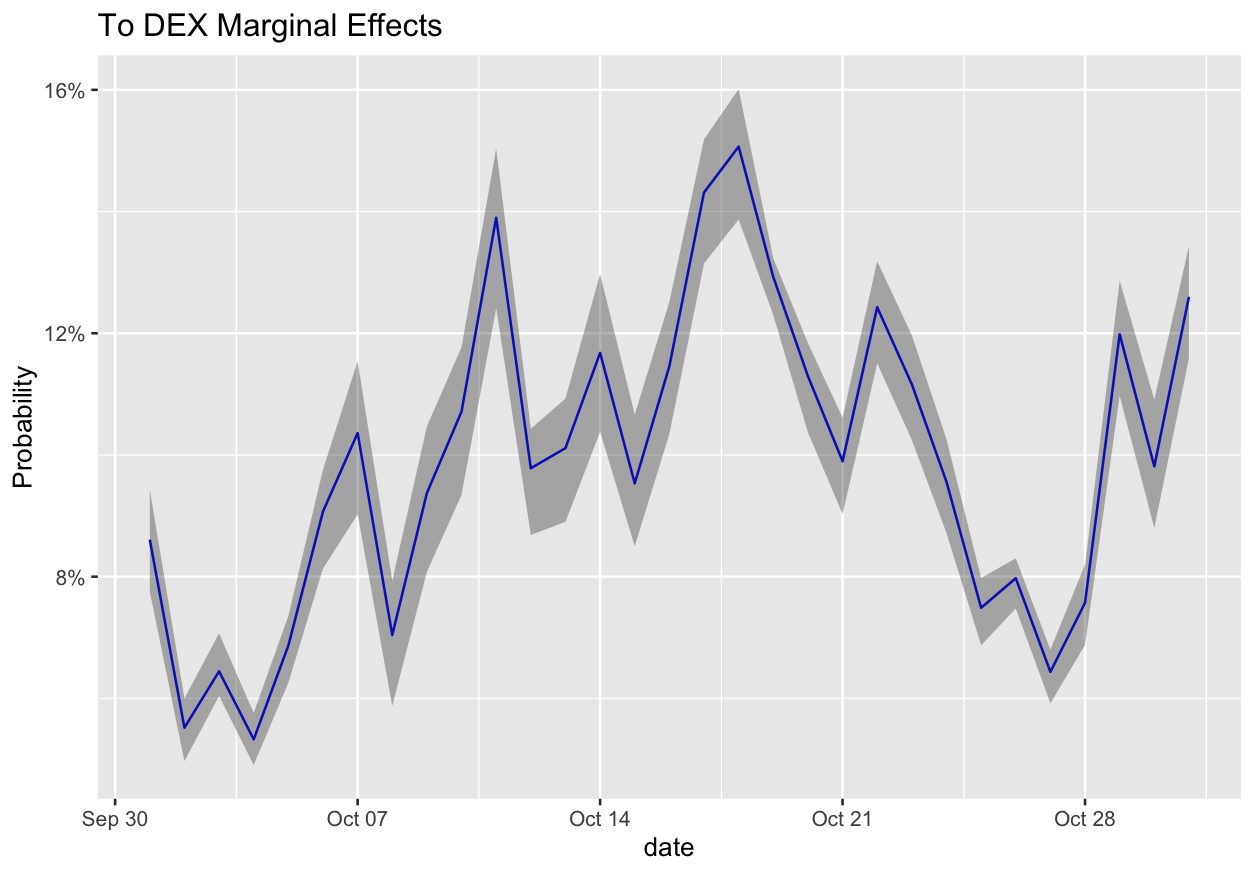} \\
         \includegraphics[width = 0.4\textwidth]{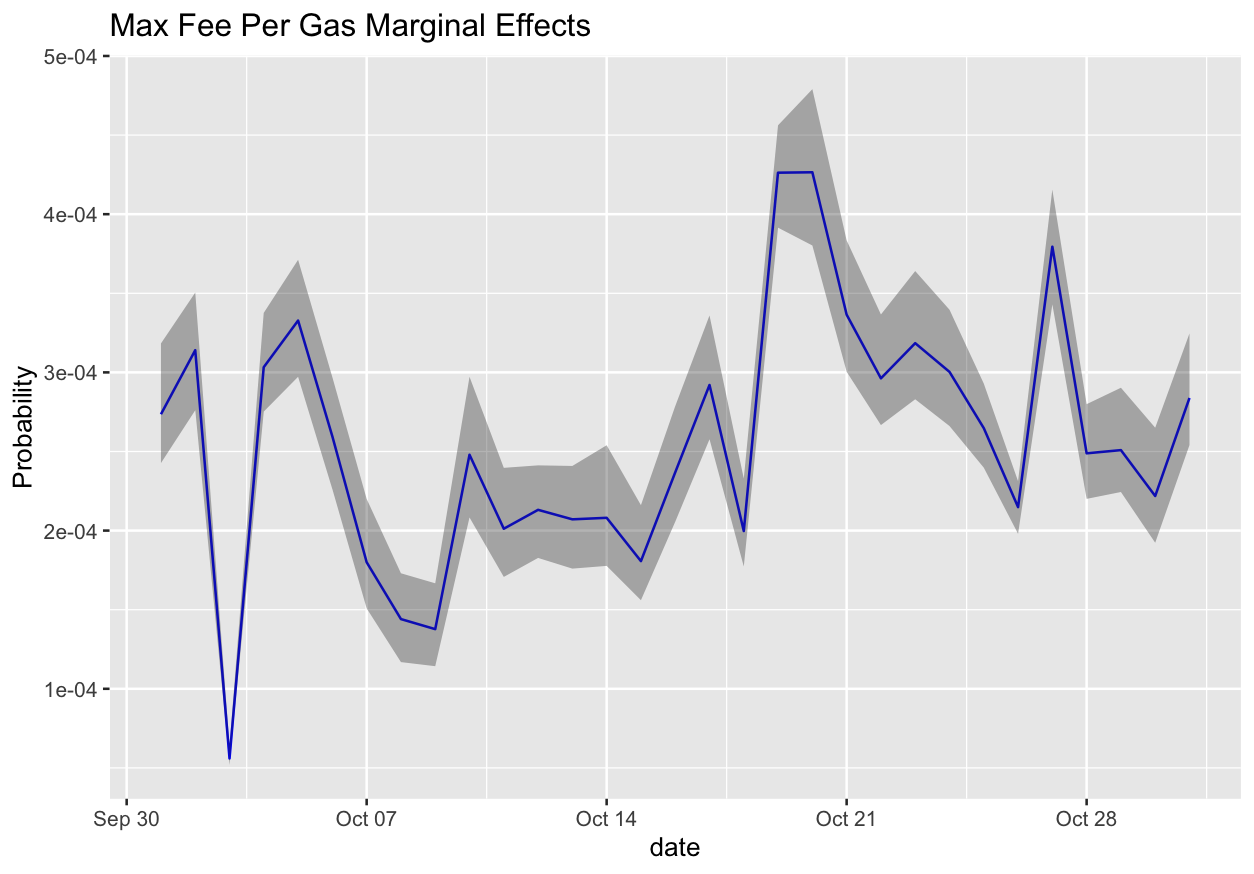} & \includegraphics[width = 0.4\textwidth]{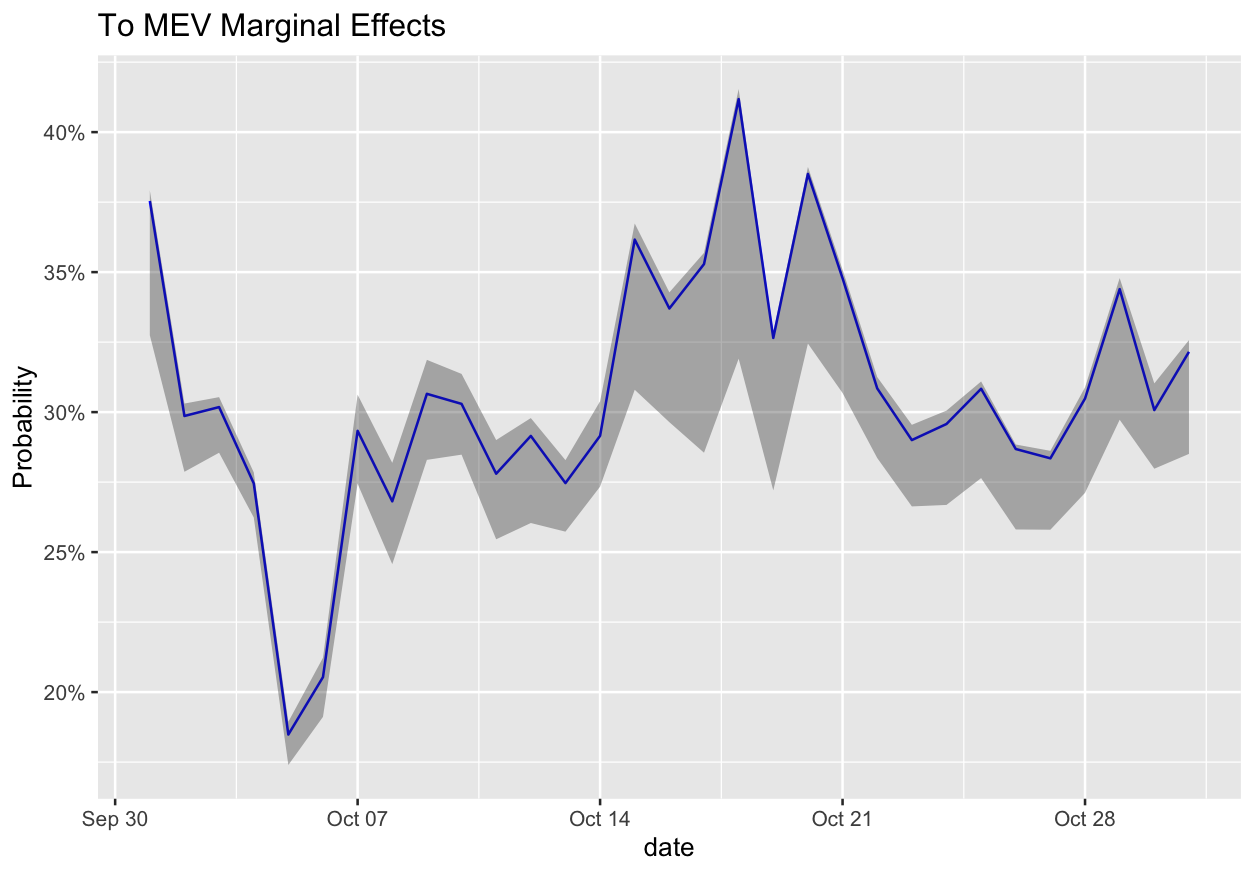} \\ 
    \end{tabular}	
		\end{center}
		\small
		\texttt{Note: } Marginal effects correspond to increased (decreased) probability of a transaction being in the first quartile of block transactions. Blue lines are median marginal effects, while the lighter bands correspond to the 10th and 90th percentiles of marginal effects. 
        \end{minipage}
    \label{fig:enter-label}
\end{figure}

The median marginal effects, as well as the marginal effects at the 10th and 90th percentiles, are shown in Figure \ref{fig:daily_quantile}. Again, we report marginal effects from the full model with controls. There is some evidence that the distribution of marginal effects is skewed and that the median may be more representative than the mean. However, results considering the average marginal effects are qualitatively similar to results based on median marginal effects. The median marginal effects for October 1st are $37.56\%$ for transactions from DEX, $8.61\%$ for transactions to DEX, $1\%$ for maximum fee per gas, and $37.54\%$ for transactions to an MEV agent. Note that as maximum fee per gas is the lone continuous variable, we impose the normalization that the marginal effect reported is for a $1\%$ change in the probability of a transaction appearing in the first block quartile. Additionally, we observe limited intraday variation of the marginal effects. 

We find that approximately 4,600 additional units of gas per transaction on average are required to prevent transaction reordering. This is equivalent to approximately \$0.20 of additional gas per transaction. This is nearly identical to the estimated effects for October 1st, \$0.21 per transaction. There were 35,799,578 transactions recorded for the month of October. Thus, the shadow price of time priority is approximately 164.8 billion units of gas or almost \$7.2 million.

\section{Sandwich Attacks} \label{sandwich}

A sandwich attack involves the insertion of a front running transaction by the block builder before another transaction in the mempool.  The sandwich is often completed by a back run transaction so that the position is completely flat and profits can be computed.  

Table \ref{tab:sandwich_example} contains an example from May 10, 2025, where an order to swap ETH for RATO on Uniswap V2 is front run by a large purchase.

\begin{table}[H]
  \centering
\begin{threeparttable}
  \caption{Sandwich Attack on May 10, 2025 in Block 22450093}
  \label{tab:sandwich_example}
  
    \begin{tabular}{rrrrr}
    \hline
    \multicolumn{1}{l}{Position in Block} & \multicolumn{1}{l}{ETH} & \multicolumn{1}{l}{Swap Direction} & \multicolumn{1}{l}{RATO} & \multicolumn{1}{l}{RATO  in ETH} \\
    \hline \hline
    10    & 1.586925 &  $-->$     & 319,495,865.86 & 4.966966E-09 \\
    11    & 0.830000 &  $-->$     & 157,358,171.48 & 5.274591E-09 \\
    12    & 1.641604 &  $<--$     & 319,495,865.86 & 5.138107E-09 \\
    \hline
    \end{tabular}%
\begin{tablenotes}
      \small
      \item \texttt{Note: } The example is from the EigenPhi \href{https://eigenphi.io/mev/ethereum/tx/0xad6bcca7a2c65f87abe7e6bf388b097a31fa9ae1ea65cf8016790ebb01b3d6fd}{dashboard}.
    \end{tablenotes}
    \end{threeparttable}%
\end{table}%

Because of the front running, the amount of RATO received for the 0.83 ETH is lower.  If the trader had not been front run, he would have received 167,104,025 RATO, almost 10,000,000 more.  The front run trade is closed in transaction 12 by selling the initial RATO purchased in transaction 10.  The 319,495,866 RATO now returns 1.641604 ETH, producing revenues of 0.054139 ETH (\$128.344498) .  EigenPhi estimates costs (gas plus DEX fees) of \$124.026572, so the net profit is only \$4.317926.  The back run transaction\footnote{Hash:0x989e2455430f20811de6682e95a7b87d2585305fb12627895a4df032f795cbe7} fee is 0.05274 ETH, so the builder actually receives the bulk of the profit.

\begin{figure}[H]
	\centering
  \centering
		\caption{Number of Sandwich Attacks During 2023-2024}
		\label{fig:daily_attack}
        \begin{minipage}{0.97\linewidth}
        \begin{center}
			\includegraphics[width=0.97\textwidth]{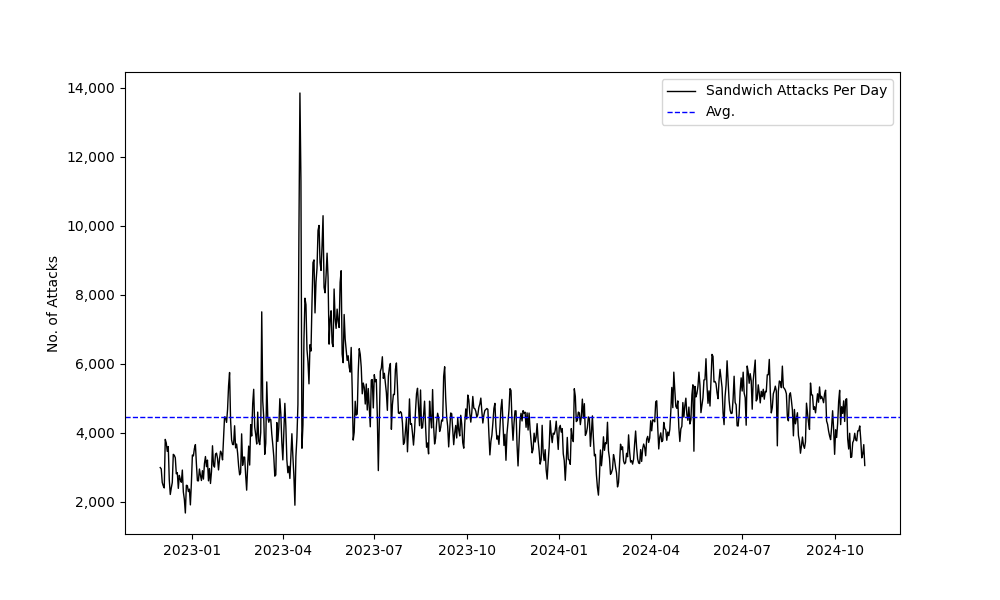} \\
		\end{center}
		\small
		\texttt{Note: } The blue line is the average number of attacks per day in the sample.
    \end{minipage}
\end{figure}

Sandwich attacks are very frequent, as shown in Figure \ref{fig:daily_attack}, averaging more than 4,400 per day in EigenPhi's sample from October 2022 to September 2024.

\subsection{Detailed Analysis of October 2024} \label{October 1st}

\begin{figure}[H]
	\centering
  \centering
		\caption{Sandwich Attacks in October 2024}
		\label{fig:sandwich_202410}
        \begin{minipage}{0.97\linewidth}
        \begin{center}
			\includegraphics[width=0.97\textwidth]{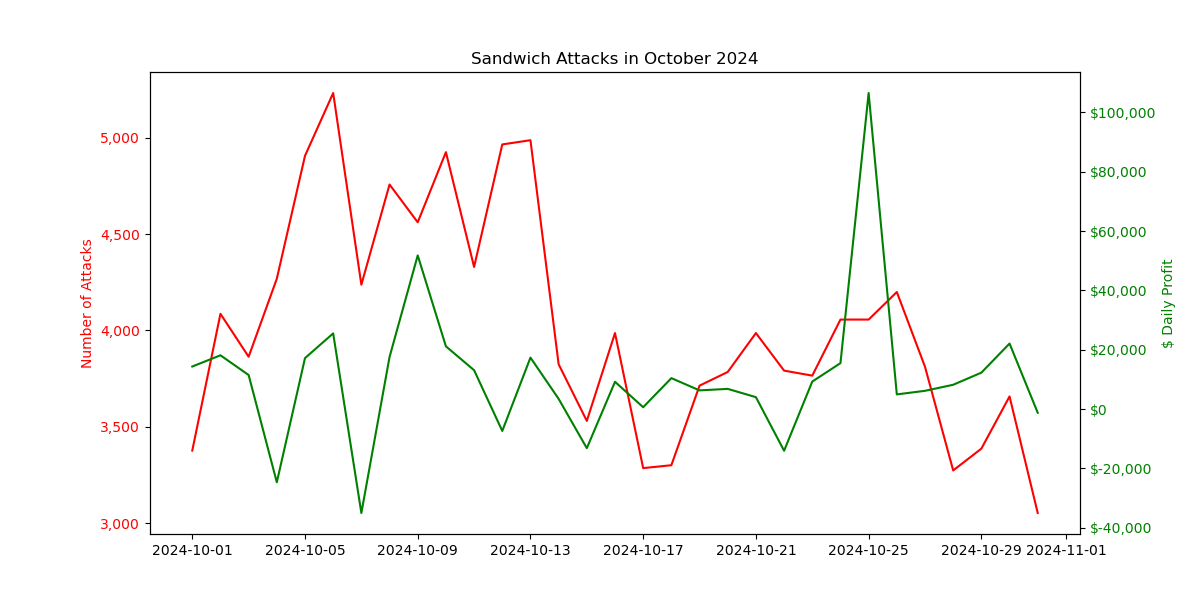} \\
		\end{center}
		\small
		\texttt{Note: } The figure shows daily estimates of the number of attacks and profits from sandwich attacks.
    \end{minipage}
\end{figure}

In this section, we focus on a more detailed analysis of individual attacks during our estimation sample from October 2024.  For each attack, we have the block position of the front run, sandwich, and back run portions of each attack.  We summarize in Figure \ref{fig:sandwich_202410} daily attacks and profits.

Our dataset identifies the portions of the sandwich by front run, the sandwiched transaction, and the back run.  One important thing we have identified is that the back run transaction, in which the attacker flattens their position and records a profit, pays substantially higher gas fees than other transactions.  We verify this using a simple t-test in Table \ref{tab:backrun_gas_test}.

\begin{table}[H]
  \centering
\begin{threeparttable}
  \caption{Gas Fees on Back Run Transactions}
  \label{tab:backrun_gas_test}
    \begin{tabular}{lrr}
\hline
          & \multicolumn{1}{l}{Avg. Gas Fees ETH} & \multicolumn{1}{l}{Std. Dev.} \\
\hline \hline
    Back run Transactions & 0.0248 & 0.0946 \\
    Non-Back run & 0.0025 & 0.0036 \\
\hline
          &       &  \\
    T-stat & 137.9423 &  \\
    P-value & 0.0000 &  \\
\hline
    \end{tabular}%
\begin{tablenotes}
      \small
      \item \texttt{Note: } The table compares the average gas fees for the back run transaction to the fees of the other two legs of the sandwich.  The sandwich positions are from EigenPhi and the gas fees are from Google Big Query.  The sample is the month of October 2024.
    \end{tablenotes}
    \end{threeparttable}%
\end{table}%

To put these gas fees into context, we first convert them to USD using Coin Gecko prices.  The average ETH price for the month of October 2024 was \$2,520.05, so the back run transaction fees averaged \$62.70.  For the 124,946 sandwich attacks in the month, we estimate total gas fees for the MEV at 7.89 million dollars.  For additional context, block builders paid an average of 0.1554 ETH for each block, so approximately 9.6\% of that was funded by gas revenues from the back run transactions.

\section{Sandwich Effects in the Baseline Model} \label{sandwich effects}

In this section, we now re-specify and estimate the model from Section \ref{block position} to analyze the gas impacts of the sandwich attacks.  We include a representative estimation of the new model for October 1st to contrast it with our previous estimate in Table \ref{tab:ordered_model_estimates_probit}.\footnote{To address the issue of numerical instability, all continuous variables were first Winsorized at the 0.1th and 99.9th percentiles before undergoing z-score normalization as in prior estimates.}

\begin{table}[H]
\centering
\begin{threeparttable}
\caption{Extended Model Estimates October 1, 2024}
\label{tab:model_estimates_sandwich}
\begin{tabular}{lrrrr}
\toprule
Variable & Value & Standard Error \\
\midrule
\textit{Main Variables} \\
max fee per gas & ${-1.436 \times 10^{-3}}^{***}$ & $3.515 \times 10^{-5}$ \\
to DEX & ${-6.589 \times 10^{-1}}^{***}$ & $1.499 \times 10^{-2}$ \\
to MEV & ${-1.960}^{***}$ & $3.053 \times 10^{-2}$ \\
from DEX & ${-1.747}^{***}$ & $8.559 \times 10^{-2}$ \\
from MEV & ${1.428}^{***}$ & $3.278 \times 10^{-2}$ \\
front run & ${-2.553}^{***}$ & $9.550 \times 10^{-2}$ \\
back run & ${-7.506 \times 10^{-1}}^{***}$ & $9.328 \times 10^{-2}$ \\
\textit{Mempool Position} \\
mempool quartile 1 & $-1.350 \times 10^{-2}$ & $1.109 \times 10^{-2}$ \\
mempool quartile 2 & $-6.176 \times 10^{-3}$ & $1.113 \times 10^{-2}$ \\
mempool quartile 3 & $-1.244 \times 10^{-2}$ & $1.113 \times 10^{-2}$ \\
\textit{Controls}\\
average price & ${5.595 \times 10^{-3}}^{***}$ & $2.164 \times 10^{-4}$\\
gas used & ${-2.982 \times 10^{-6}}^{***}$ & $4.567 \times 10^{-8}$\\
validator payment & ${1.272 \times 10^{-3}}^{***}$ & $2.755 \times 10^{-4}$\\
priority fee & ${-9.296 \times 10^{-2}}^{***}$ & $8.085 \times 10^{-4}$\\
sandwich cost & ${-6.095 \times 10^{-2}}^{***}$ & $3.224 \times 10^{-3}$\\
Boundary 1|2 & ${-2.901 \times 10^{-1}}^{***}$ & $1.234 \times 10^{-2}$ \\
Boundary 2|3 & ${8.308 \times 10^{-1}}^{***}$ & $1.244 \times 10^{-2}$ \\
Boundary 3|4 & $1.768^{***}$ & $1.281 \times 10^{-2}$ \\
\bottomrule
\end{tabular}
\begin{tablenotes}
\small
\item \textit{Note:} Negative coefficients indicate a higher likelihood of a transaction being placed earlier in the block. $***$ indicates significance at the 0.1\% level. 
\end{tablenotes}
\end{threeparttable}
\end{table}

Front run is an indicator variable taking the value one when a transaction is the front run transaction in a sandwich attack. Back run is similarly defined, taking the value one when a transaction is the back run transaction in a sandwich attack. We note that the coefficient estimates do not vary significantly from the estimates in the more limited model. Importantly, both the front run and the back run indicators are highly statistically significant and economically significant. As further control variables, we add the maximum priority fee and sandwich cost. The maximum priority fee separates the priority fee portion from the overall maximum fee allowed by the transaction originator. The sandwich cost measures the transaction size of the sandwiched transaction. 

The average marginal effects for the main variables of interest are presented in Table \ref{tab:avg_marginal_sandwich}. As with prior estimation results, the model considered is the model with the full set of controls.

\begin{table}[H]
  \centering
  \begin{threeparttable}
  \caption{Average Marginal Effects for October 1, 2024}
    \begin{tabular}{lrrr}
    \toprule
          & & Marginal Effects        &  \\
    Variable & \multicolumn{1}{l}{Prob. In First Quartile} & \multicolumn{1}{l}{Gas Equivalent Cost} & \multicolumn{1}{l}{Cost in USD} \\
    \midrule
    \midrule
    $\text{max fee per gas}^*$ & 1\% & 40 & \$0.0023 \\
    to DEX & 12.68\% & 499 & \$0.0291 \\
    to MEV & 39.13\% & 1,540 & \$0.0898 \\
    from DEX & 34.89\% & 1,373 & \$0.0801 \\
    from MEV & -19.09\% & 752 & \$0.0438 \\
    front run & 47.96\% & 1,888 & \$0.1100 \\
    back run & 14.64\% & 576 & \$0.0336 \\
    
    \bottomrule
    \end{tabular}%
  \label{tab:avg_marginal_sandwich}%
     \begin{tablenotes}
      \small
      \item \texttt{Note: } $*$ denotes that the variable max fee per gas is continuous and thus we normalize effects to be equivalent to a 1\% change in the probability of appearing in the first block quartile. All other variables are discrete and thus marginal effects reflect the change in probability when the value of the variable goes from 0 to 1. The table provides estimates of the marginal effects as the change in the probability of the transaction being included in the first quartile. We measure the harm to the participant by measuring the gas and dollar cost of achieving a similar block position.  We use average daily gas costs of 22.45 Gwei and the closing Ethereum price of \$2,597.34.
    \end{tablenotes}
    \end{threeparttable}%
\end{table}

The front run and back run transactions of the sandwich attack both show large marginal effects of 47.96\% and 14.64\%, respectively. The back run marginal effect is similar in magnitude to the marginal effects previously observed, but the front run marginal effect is the largest marginal effect by some margin. A front run transaction is nearly 50\% more likely to appear in the first quartile of block transactions, and being the front run transaction in a sandwich attack is the single most impactful transaction characteristic for predicting a transaction's position in the finalized block. The gas costs associated with front run transactions are also much larger than any of the marginal effects we saw in the baseline model from Table \ref{tab:avg_marginal_probit}. Front run marginal effects are approximately \$0.11 per transaction.

\begin{figure}[H]
    \centering
    \caption{Daily Average Marginal Effects Sandwich Transactions}
    \label{fig:daily_sandwich}
        \begin{minipage}{0.97\linewidth}
        \begin{center}
		  \begin{tabular}{cc}
         \includegraphics[width = 0.4\textwidth]{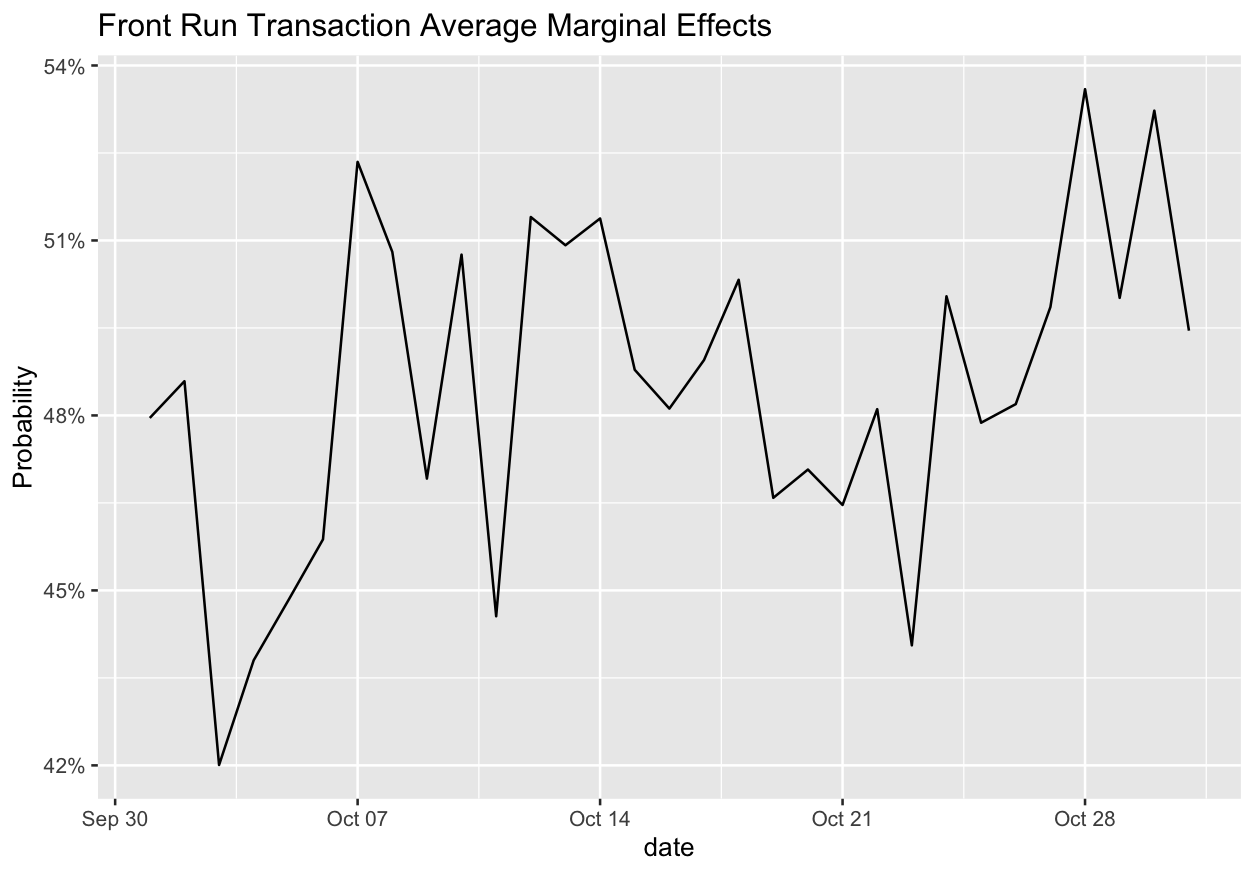} & \includegraphics[width = 0.4\textwidth]{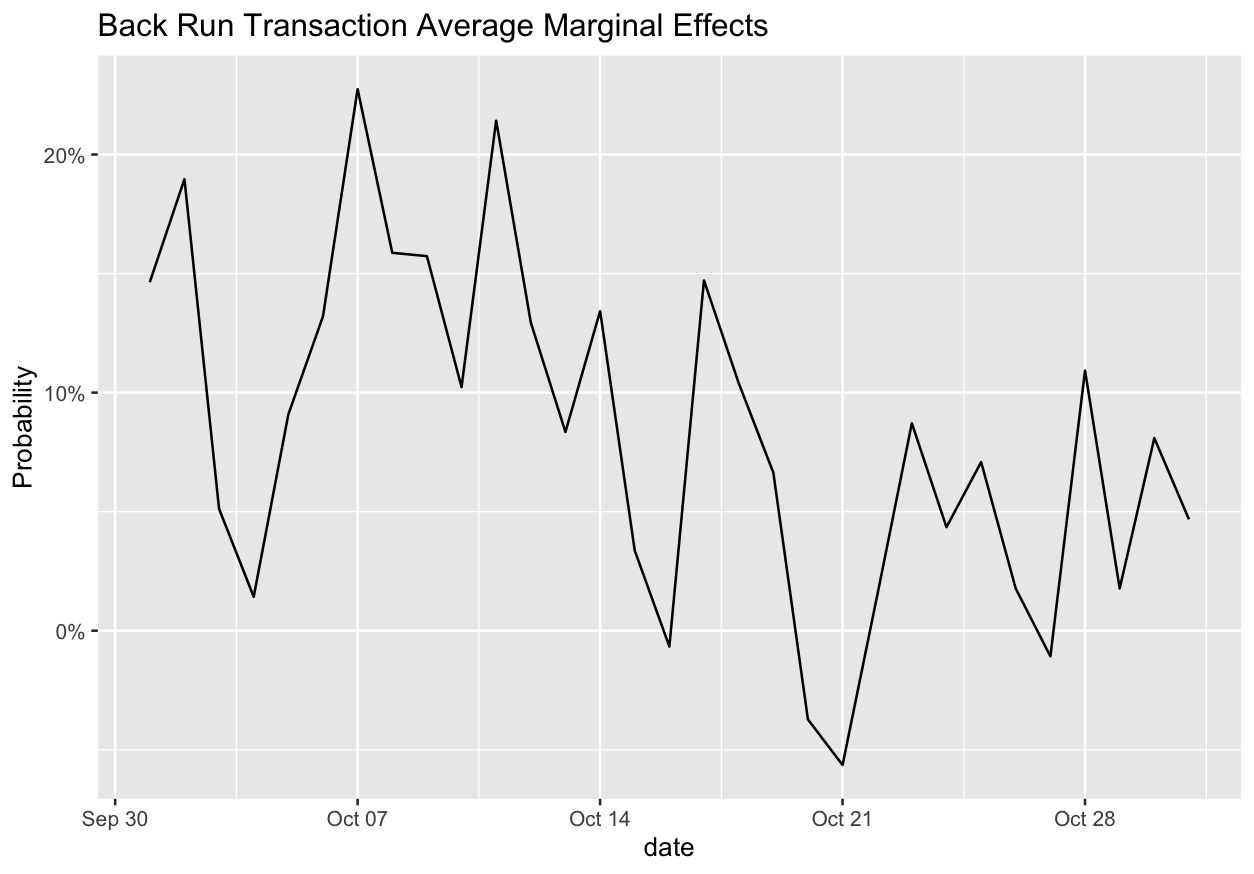} \\
    \end{tabular}	
		\end{center}
		\small
		\texttt{Note: } Average marginal effects correspond to increased (decreased) probability of a transaction being in the first quartile of block transactions
        \end{minipage}
    \label{fig:enter-label}
\end{figure}

The daily average marginal effects for the indicator variables corresponding to a transaction being the front run transaction in a sandwich attack or the back run transaction in a sandwich attack for October 2024 are shown in Figure \ref{fig:daily_sandwich}. The marginal effects for both indicator variables appear relatively stable throughout the sample period, although it should be noted that there is a limited downward trend in the marginal effects of back run transactions. This generally matches the trend of more frequent sandwich attacks at the beginning of the month and less frequent attacks closer to the end of the month, as seen in Figure \ref{fig:sandwich_202410}. 

Extending our model to the entire sample, our results remain similar to our estimates using data from only October 1st. The average marginal effect for front run transactions in October 2024 is approximately 0.4848; keeping all other transaction characteristics constant, a transaction that is the front run transaction in a sandwich attack is approximately 48.48\% more likely to appear in the first block quartile. The marginal effects for back run transactions are smaller than those for front run transactions but they are still significant, averaging 8.26\% for October 2024. 

There were 124,946 sandwich attacks in October 2024. We average the average marginal effects across each day in our sample  and find average marginal effects of \$0.1118 and \$0.0260 for front run and back run transactions, respectively. Incorporating the average marginal effects in USD of both the front and back run transactions, \$0.1118 + \$0.0260, we estimate costs to participants of \$17,227.56 just from reordering.  Please note that this does not include the costs of the bad executions from being sandwiched.

The losses faced by market participants due to poor transaction execution are a highly-related but distinct risk faced by investors. We do not attempt to directly calculate these losses. \citet{Capponi_JFE} estimate an average monthly loss of 4,500 Ether due to sandwich attacks in their sample period spanning May 2020 to September 2022. Extrapolating to our sample period of October 2024, this translates to a loss of approximately \$11.34 million. 

Our measure of aggregate losses is distinct from the idea of reordering slippage developed in \citet{Adams_Slippage}. Reordering slippage measures losses in terms of differences between the quoted price and the execution price of swaps. Adversarial slippage measures reordering losses due to the insertion of DEX transactions not previously in the mempool. Our analysis focuses more broadly on losses due to transaction reordering for all transaction types.

EigenPhi has made an estimate of the net profits from sandwiches that incorporates gas and DEX fees. For the sample period November 30, 2022  to October 31, 2024, estimated net profits are \$81,079,748.36. For the month of October 2024, net profits are estimated at \$337,279.33. While sandwich attacks are arguably the most egregious form of front running, the issue of transaction reordering is much broader. Comparing our estimate of aggregate losses, \$7.18 million, to the direct effect of sandwich transaction reordering, we find that \$7.16 million or 99.76\% of losses due to reordering are not directly due to transaction reordering associated with sandwich attacks.

\begin{figure}[H]
    \centering
    \caption{Daily Quantile Marginal Effects Sandwich Transactions}
    \label{fig:daily_sandwich_quantile}
        \begin{minipage}{0.97\linewidth}
        \begin{center}
		  \begin{tabular}{cc}
         \includegraphics[width = 0.4\textwidth]{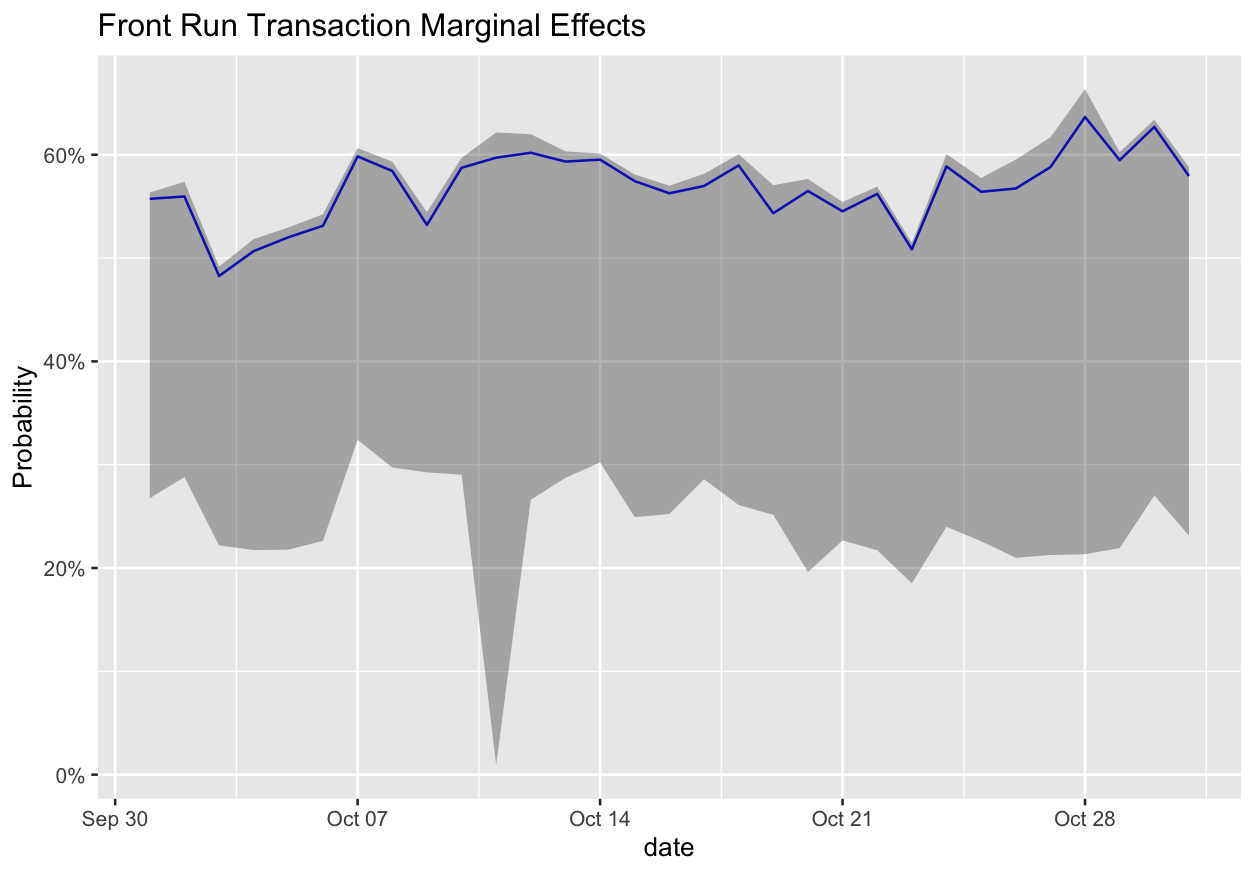} & \includegraphics[width = 0.4\textwidth]{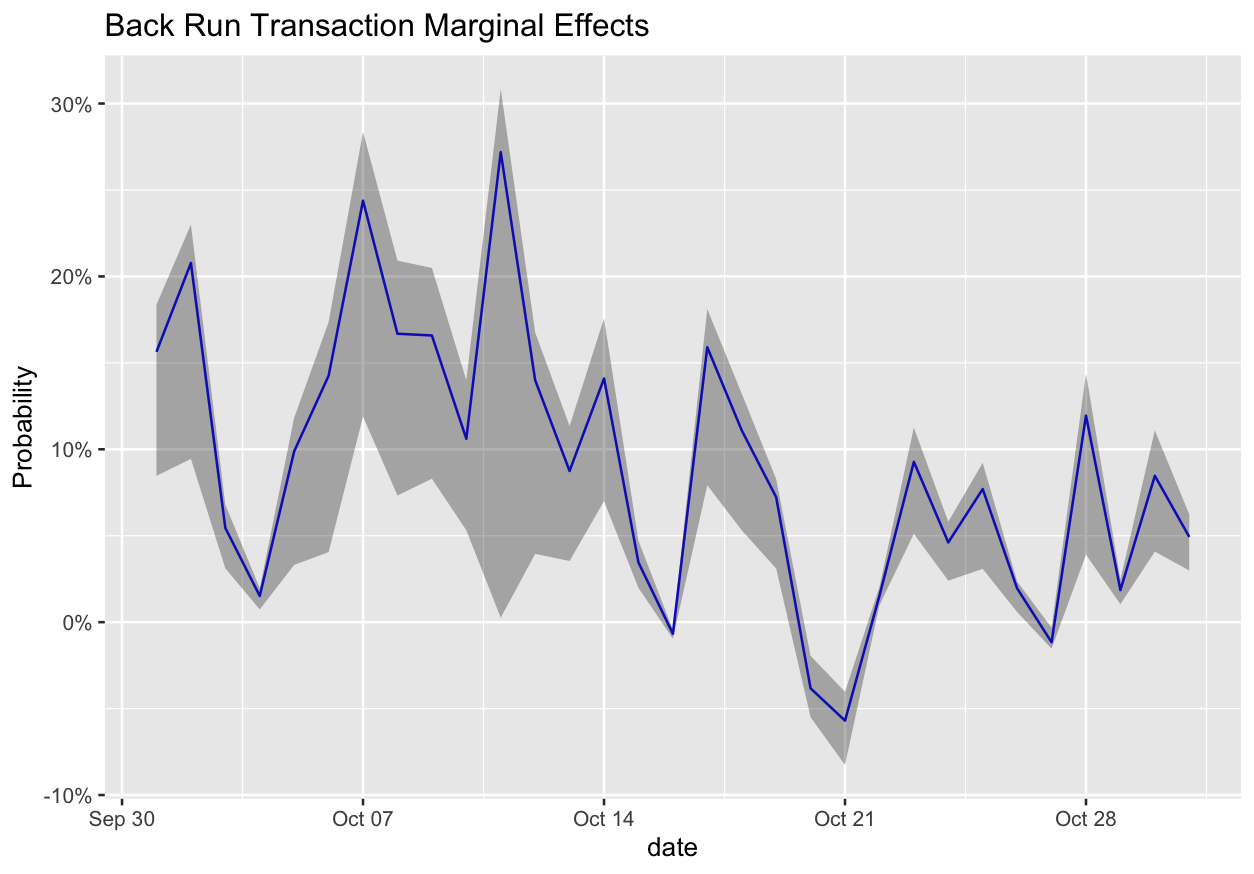} \\
    \end{tabular}	
		\end{center}
		\small
		\texttt{Note: } Marginal effects correspond to increased (decreased) probability of a transaction being in the first quartile of block transactions. Blue lines are median marginal effects, while the lighter bands correspond to the 10th and 90th percentiles of marginal effects.
        \end{minipage}
    \label{fig:enter-label}
\end{figure}

The median marginal effects along with the marginal effects at the 10th percentile and the 90th percentile for the front run and back run transactions are shown in Figure \ref{fig:daily_sandwich_quantile}. The marginal effects for front run transactions show more intra-day variation compared to the marginal effects in the non-sandwich model in Figure \ref{fig:daily_quantile}. We also see stronger evidence that the distribution of marginal effects is heavily left-skewed.  

We calculate dollar-equivalent marginal effects at the median for front run and back run transactions as follows:

\begin{multline}\label{Eq: 4}
    \text{Marginal Effect} = \frac{\partial F / \partial {x_i}}{\partial F / \partial \text{max fee per gas}} \times \text{average daily gas price} \\ \times \text{average daily Ethereum price}.
\end{multline}

Equivalently, we compare the median marginal effect of front run or back run transactions to the median marginal effect of maximum fee per gas and then convert to an equivalent dollar cost.  The median of the median dollar-equivalent marginal effects is \$0.0152 and \$0.0897 for back run and front run transactions, respectively. Aggregating as before, this suggests that the median reordering cost directly attributable to sandwich transactions is \$13,106.84. This small aggregate reflects the fact that sandwich attacks still comprise a small portion of the total number of transactions.

\section{Conclusion} \label{conclusion}

Equity markets provide a useful lens for considering MEV. Although those regulated markets have enshrined price-time priority, they also provide mechanisms for consolidating the market and routing to protected quotes. The equity markets have sanctioned expenditure on trading technology that might give an edge to institutional investors.  The mempool is not centralized and would require protocol reform  to impose time priority within a decentralized ledger.

 We estimate the marginal effects of MEV transaction reordering and find that in our baseline model, investors would pay nearly \$7.2 million to ensure that they remained in the first quartile of the block. We believe this is the first estimate of this cost in the literature. This is of the same order of magnitude as prior estimates of sandwich losses and losses due to reordering slippage.\footnote{\citet{ReorderSlippage}.} 
 
 Sandwich attacks, in which a transaction is front run, are extremely frequent, averaging more than one every two blocks. Even if validators share in the extractable value, this is of little solace to DEX traders who are front run. 
 
 Gas fees on sandwich transactions subsidize 9.6\% of the MEV payment for the block. These attacks, although infrequent, are associated with significant transaction reordering. The corresponding marginal effects are 34\% larger than any of the other marginal effects observed. Being the front run transaction in a sandwich attack is the single most important transaction characteristic for predicting a transaction's final block position. 

Transaction reordering threatens the integrity of the online swap platforms. Base has prioritized both time and gas fees in their Layer 2, and it is currently the most active among the Layer 2's in DEX transactions.\footnote{\url{https://www.galaxy.com/insights/research/base-l2-thriving-on-priority-fees-and-dex-activity}} Uniswap has built a Layer 2 network called Unichain\footnote{\url{https://blog.uniswap.org/rollup-boost-is-live-on-unichain}} to ``build blocks inside a trusted execution environment (TEE).''  The TEE orders transactions based purely on priority fees.  

Despite the theoretical appeal and rapid early adoption of such technology, by March 2026, it appears that adoption remains limited at best. Unichain grew rapidly between its launch in February 2025 to its peak in July 2025, growing from \$346.92 million to \$14.304 billion in monthly DEX volume during this period. However, after reaching this peak, adoption dropped rapidly, reaching just \$394.58 million of monthly DEX volume in March 2026.\footnote{\url{https://defillama.com/chain/unichain?chainFees=false&dexsVolume=true}}. 

An alternative way to avoid sandwich attacks and MEV extraction more broadly is to execute transactions faster. Base executes transactions every 2 seconds compared to every 12 seconds on the Ethereum network. Base, has fared better in terms of adoption metrics, but has similarly fallen from its peak volume. They reached an all time high monthly DEX volume of \$53.109 billion in October 2025, yet by March 2026 their monthly DEX volume fell to just \$24.355 billion.\footnote{\url{https://defillama.com/dexs/chain/base}}

\citet{MEV_mitigation} provide a survey of MEV mitigation techniques, with a primary emphasis on alternative sequencing approaches. They recognize the value of a single time-priority-based sequencer such as those deployed by Arbitrum and Optimism, but express concern about the network having a single point of failure among other concerns.

A further solution that has been proposed is the use of private transaction pools, as discussed in \citet{Capponi_JFE}. Private transaction pools reduce the risk of front running as transactions are sent directly from market participants to block builders, never being exposed to the public mempool. However, private pool users face execution risk and higher transaction fees than users of the public mempool, and private pool adoption remains limited. 

An additional potential solution to the MEV  problem is the encryption of pending transactions until the blocks have been finalized. The precise content and order of pending transactions is vital for a successful front running attack, so by keeping this information encrypted until the blocks have been finalized, front running is effectively mitigated. A specific implementation of this is called batched threshold encryption, which \citet{batch_encryption} describe as follows. ``[A] batch of encrypted transactions is selected by a committee and only decrypted after block finalization." However, this proposed solution is relatively slow and works against the Ethereum Foundation's desire to speed up transactions.   

We aim to highlight the severity of the MEV problem and to motivate meaningful action to address it.  If blockchains want to compete with traditional financial institutions like equity markets, the MEV problem must be mitigated.

\pagebreak
\begingroup
\setstretch{1.25}
\printbibliography

@article{shkilko_micro,
author={Shkilko, Andriy and Sokolov, Koonstantin},
title={{Every cloud has a silver lining: Fast trading, microwave connectivity and trading costs}},
journal={Journal of Finance},
volume={75}, 
pages={2899-2927},
year={2020}
}

@misc{monnot_auctions,
author={Monnot, Barnabe},
title={{More pictures about proposers and builders}},
url = {https://mirror.xyz/barnabe.eth/QJ6W0mmyOwjec-2zuH6lZb0iEI2aYFB9gE-LHWIMzjQ},
year = {2024}
}

@misc{zhu_revert,
author={Zhu, Brian and Wan, Xin, Moallemi and Ciamac and Robinson, Dan and Bachu, Brad},
title={{Quantifying the Value of Revert Protection}},
url = {https://arxiv.org/pdf/2410.19106},
year = {2025}
}

@article{KapengutPoS,
author={Kapengut, Eli and Mizrach, Bruce},
title={{An Event Study of The Ethereum Transition to Proof-of-Stake}},
journal={Commodities},
volume={2},
pages={96-110},
year={2023},
}

@article{BattalioNYSE,
author={Battalio,  Robert and Jennings,Robert, and  McDonald, Bill},
title={{Deviations from time priority on the NYSE}},
journal={Journal of Financial Markets},
volume={53},
pages={1-17},
year={2021},
}

@misc{ButerinPBS,
author={Vladimir Buterin},
title={{Proposer/block builder separation-friendly fee market designs}},
url={{https://ethresear.ch/t/proposer-block-builder-separation-friendly-fee-market-designs/9725}},
year={2021},
}

@article{EskandariFront,
author={Eskandari, S. and Moosavi, S. and Clark, J.},
title={{Sok: Transparent dishonesty: front-running attacks on blockchain}},
journal={International Conference on Financial Cryptography and Data Security},
pages={170–189},
year={2019},
}

@article{CapponiFrontRun,
author={Capponi, A. and Jia, R. and Wang, Y.},
title={{Allocative Inefficiencies in Public Distributed Ledgers}},
journal = {SSRN Electronic Journal},
url={https://papers.ssrn.com/sol3/papers.cfm?abstract_id=3997796},
year={2022},
}

@misc{ReorderSlippage,
author = {Kim, Evan and Akdeniz, Murat},
title = {{Trading in the Dark: Measuring Reordering Slippage in MEV}},
url = {https://frontier.tech/measuring-reordering-slippage-in-mev},
year = {2024}
}

@article{Capponi_JFE,
    author = {Capponi, Agostino and Jia, Ruizhe and Wang, Kanye Ye},
    title = {Maximal Extractable Value and Allocative Inefficiencies in Public Blockchains},
    journal = {Journal of Financial Economics},
    volume = {172},
    pages = {104132},
    year = {2025}
}

@article{Wooldridge_2014,
    author = {Wooldridge, Jeffrey M.},
    title = {Quasi-maximum likelihood estimation and testing for nonlinear models with endogenous explanatory variables},
    journal = {Journal of Econometrics},
    volume = {182},
    pages = {226 -- 234},
    issue = {1},
    year = {2014}
}

@article{Burian_MEV,
    author = {Burian, Jonah and Crapis, Davide and Saleh, Fahad},
    title = {MEV Capture and Decentralization in Execution Tickets},
    journal = {Working Paper},
    year = {2024}
}

@article{euro_Latency,
    author = {Rzayev, Khaladdin and Ibikunle, Gbenga and Steffen, Tom},
    title = {The market quality implications of speed in cross-platform trading: Evidence from Frankfurt-London microwave networks},
    journal = {Journal of Financial Markets},
    issue = {66},
    pages = {100853},
    year = {2023} 
}

@article{batch_encryption,
    author = {Bormet, Jan and Faust, Sebastian and Othman, Hussein and Qu, Ziyan},
    title = {BEAT-MEV: Epochless Approach to Batched Threshold Encryption for MEV Prevention},
    journal = {34th USENIX Security Symposium},
    year = {2025}
}

@misc{MEV_mitigation,
    author = {Alipanahloo, Zeinab and Senhaji Hafid, Abdelhakim and Zhang, Kaiwen},
    title = {Maximal Extractable Value Mitigation Approaches in Ethereum and Layer-2 Chains: A Comprehensive Survey},
    url = {https://arxiv.org/abs/2407.19572},
    year = {2024}
}

@article{Adams_Slippage,
    author = {Adams, Austin and Markovich, Sarit and Chan, Benjamin Y and Wan, Xin },
    title = {Don't Let MEV Slip: The Costs of Swapping on the Uniswap Protocol},
    journal = {arXiv:2309.13648},
    year = {2025} 
}
\endgroup

\end{document}